\documentclass[siggraph,nonacm]{acmart}

\setcopyright{none}

\acmConference[arXiv]{arXiv}{Nowhere}

\usepackage{multirow}
\usepackage{multicol}
\usepackage{bm}
\usepackage{array}
\usepackage{subcaption}

\makeatletter
\newcommand\footnoteref[1]{\protected@xdef\@thefnmark{\ref{#1}}\@footnotemark}
\makeatother

\newcommand{\ra}[1]{\renewcommand{\arraystretch}{#1}}

\definecolor{nicered}{rgb}{0.75,0.1,0.2}
\definecolor{nicegreen}{rgb}{0.1,0.65,0.15}
\definecolor{niceblue}{rgb}{0.35,0.6,0.85}

\usepackage{todonotes}

\DeclareMathOperator*{\mse}{MSE} %

\begin{document}
\title{Deep Direct Volume Rendering: \\
    Learning Visual Feature Mappings From Exemplary Images
}


\begin{teaserfigure}
  \centering
        \hfill
        \subfloat[\label{fig:Teaser_input}]{\includegraphics[trim={0 0 0 0},clip,width=0.25\columnwidth]{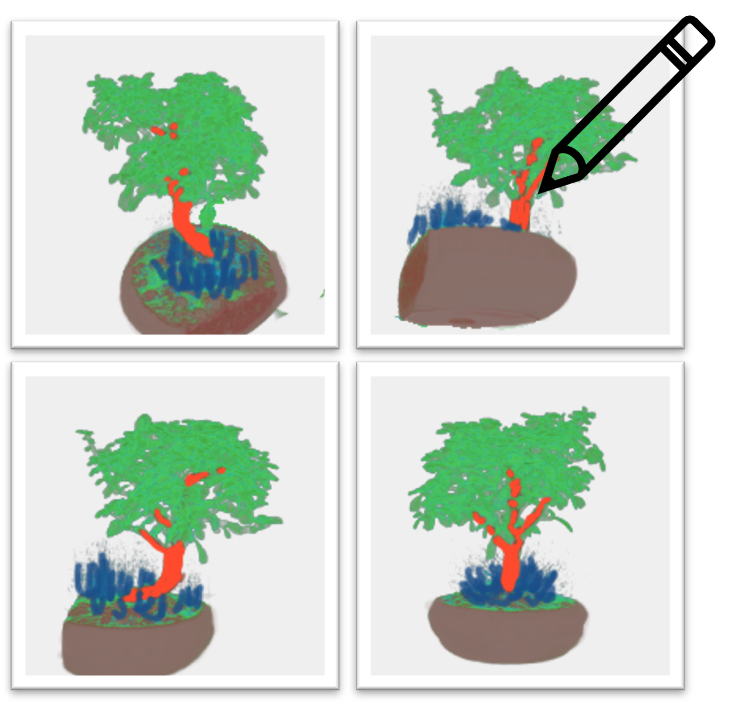}}\hfill
        \subfloat[\label{fig:teaser_pipeline}]{\includegraphics[trim={0 0 0 0},clip,width=0.315\columnwidth]{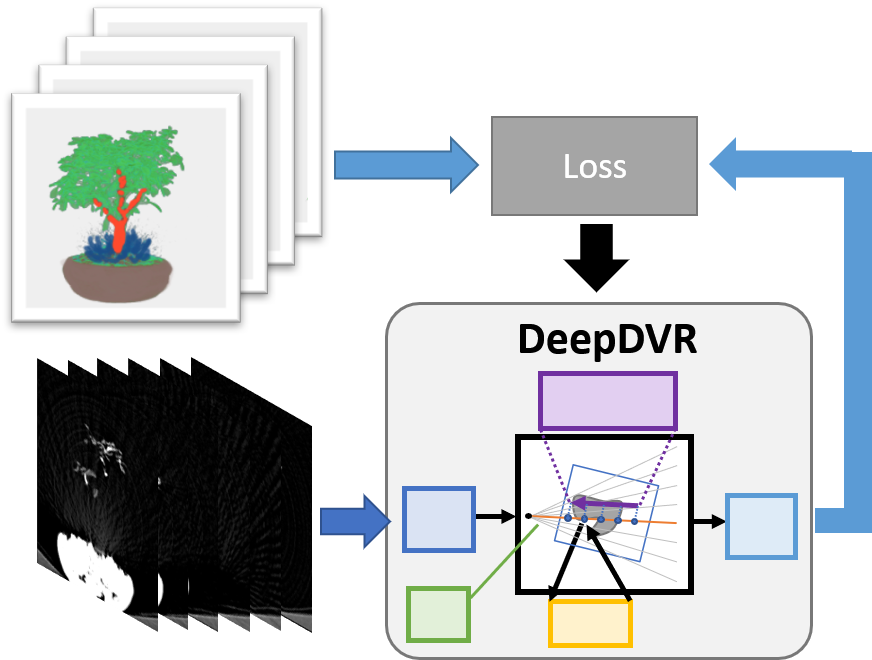}}\hfill
        \subfloat[\label{fig:Teaser_output}]{\includegraphics[trim={0 0 0 0},clip,width=0.25\columnwidth]{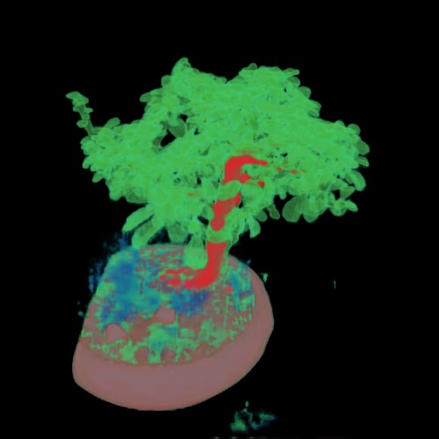}}\hfill                \hfill
    \caption{
        We train deep neural networks that explicitly model the feature extraction, classification/visual mapping and compositing of classic DVR for scientific volume rendering.
        (a) Our models are trained end-to-end from images that can, for example, be created by domain experts who can now directly specify the desired appearance in view space. 
        (b) Our DeepDVR architectures model a generalized DVR pipeline and can efficiently learn the visual features required to reproduce these images. 
        (c) The trained models easily generalize to novel viewpoints, can process volumes of arbitrary dimensions and create high resolution rendered images. 
        }
    \label{fig:Teaser}
\end{teaserfigure}




\author{Jakob Weiss}
\email{jakob.weiss@tum.de}
\orcid{0000-0002-4058-2485}
\author{Nassir Navab}
\authornotemark[0]
\email{nassir.navab@tum.de}
\affiliation{%
  \institution{Technical University of Munich}
  \streetaddress{Boltzmannstr. 3}
  \city{Munich}
  \state{Bavaria}
  \country{Germany}
  \postcode{81373}
}








\renewcommand{\shortauthors}{Weiss and Navab}

\begin{abstract}
Volume Rendering is an important technique for visualizing three-dimensional scalar data grids and is commonly employed for scientific and medical image data. Direct Volume Rendering (DVR) is a well established and efficient rendering algorithm for volumetric data.
Neural rendering uses deep neural networks to solve inverse rendering tasks and applies techniques similar to DVR.
However, it has not been demonstrated successfully for the rendering of scientific volume data.

In this work, we introduce Deep Direct Volume Rendering (DeepDVR), a generalization of DVR that allows for the integration of deep neural networks into the DVR algorithm.
\emph{We conceptualize the rendering in a latent color space, thus enabling the use of deep architectures to learn implicit mappings for feature extraction and classification, replacing explicit feature design and hand-crafted transfer functions.}
Our generalization serves to derive novel volume rendering architectures that can be trained end-to-end directly from examples in image space, obviating the need to manually define and fine-tune multidimensional transfer functions while providing superior classification strength.
We further introduce a novel \emph{stepsize annealing} scheme to accelerate the training of DeepDVR models and validate its effectiveness in a set of experiments.
We validate our architectures on two example use cases:
(1) learning an optimized rendering from manually adjusted reference images for a single volume and 
(2) learning advanced visualization concepts like shading and semantic colorization that generalize to unseen volume data.

We find that deep volume rendering architectures with explicit modeling of the DVR pipeline effectively enable end-to-end learning of scientific volume rendering tasks from target images. 
    



\end{abstract}

%
%

\ccsdesc[500]{Human-centered computing~Scientific visualization}
\ccsdesc[500]{Computing methodologies~Neural networks}
\ccsdesc[300]{Computing methodologies~Rendering}
\ccsdesc[100]{Computing methodologies~Non-photorealistic rendering}
\ccsdesc[300]{Computing methodologies~Image processing}

%
%

\keywords{volume rendering, differentiable rendering, neural rendering}

\maketitle

            \section{Introduction}

Volume rendering has become an important part of scientific visual computing as an alternative to viewing cross-sections or extracting explicit surfaces from scalar 3D datasets. 
As a widely researched technique for visualizing structural information in volumetric data sets, many of its aspects have been explored intensively in the last years.
It has been used in many application areas involving volumetric data, such as weather simulation, structural imaging in engineering, and medical imaging.

\begin{figure}
    \centering
    \includegraphics[width=\columnwidth]{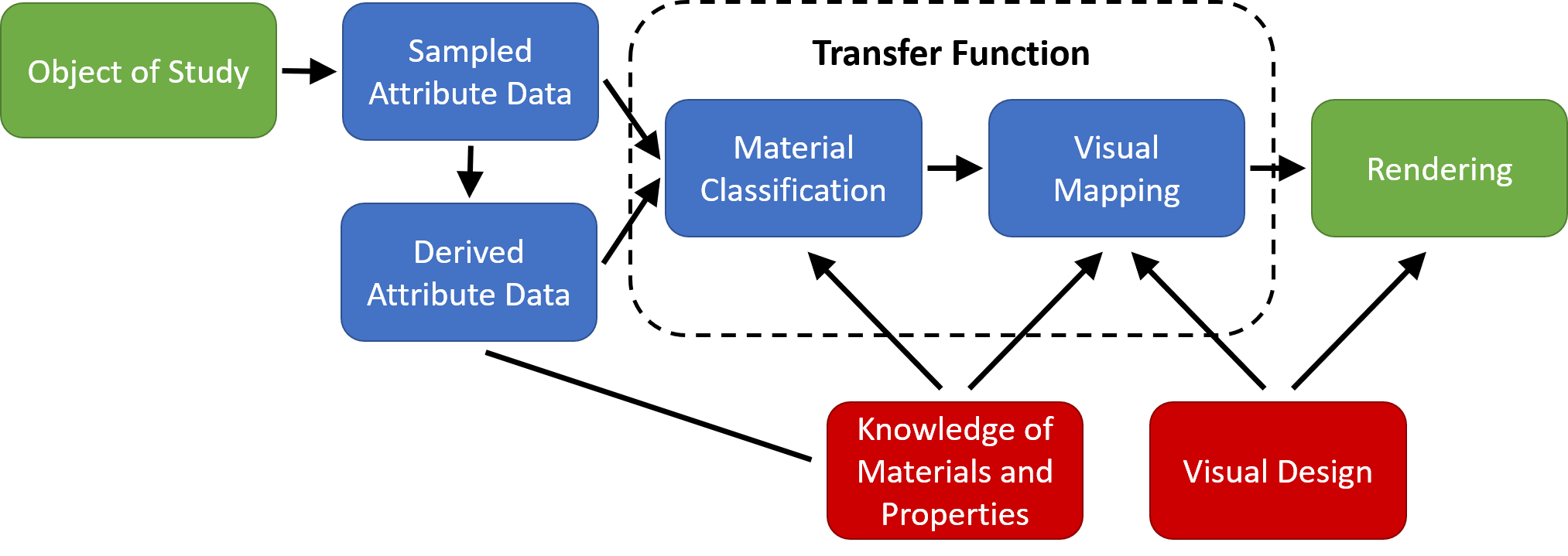}
    \caption{Schematic workflow of classic transfer function design. TF specification requires understanding of the input and derived attributes, internal parameters and the intended visual outcome. Figure adapted from Ljung et al.~\cite{Ljung2016}.}
    \label{fig:classic_tf_workflow}
\end{figure}

\begin{figure}
    \centering
    \includegraphics[width=\columnwidth]{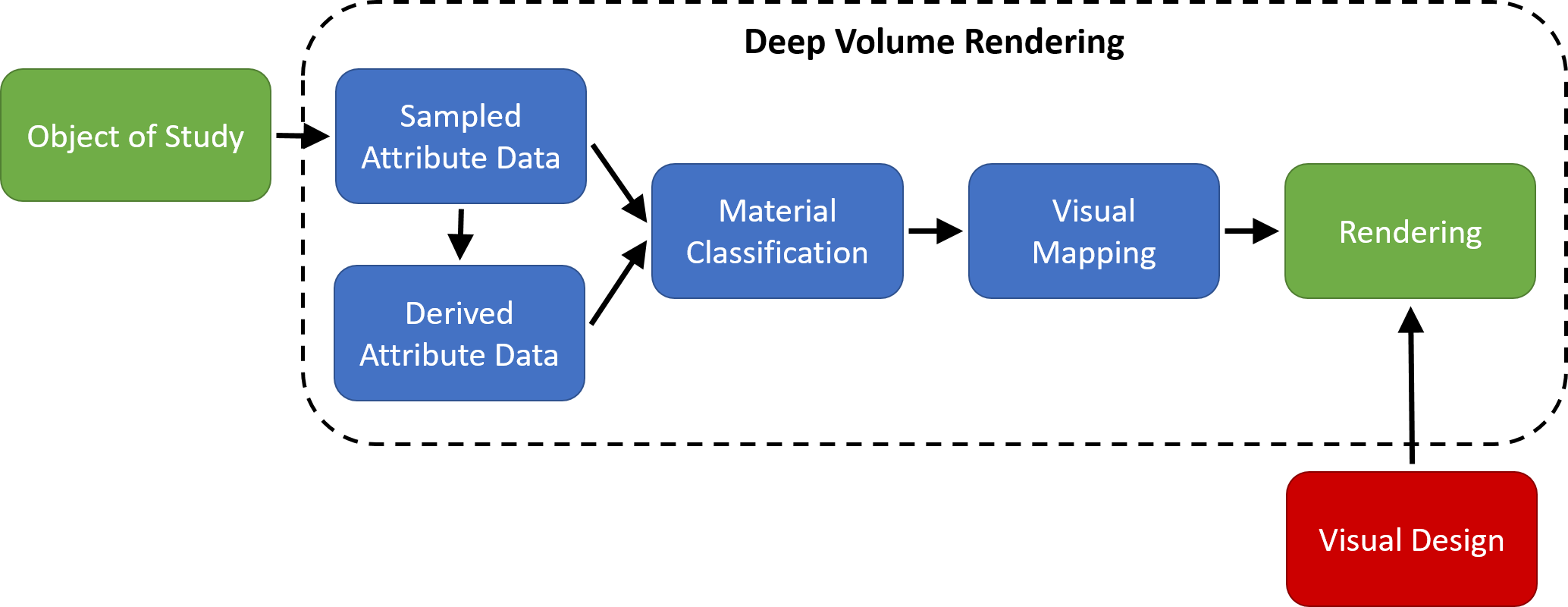}
    \caption{Modified workflow with deep direct volume rendering. Interaction is in image space and only requires specification of the desired visual outcome with less knowledge of the input attributes and materials.}
    \label{fig:DeepDVR_workflow}
\end{figure}

Volume Rendering owes much of its versatility to the use of a transfer function (TF), introduced in the earliest work on Direct Volume Rendering (DVR) \cite{Max1995} as a way of assigning optical properties (color and absorption) to the voxels of a volume.
Designing a TFs is split into a classification step and a visual mapping step (c.f. Fig.~\ref{fig:classic_tf_workflow}). The classification step derives a semantic based on original and derived voxel attributes and the visual mapping associates visual properties (typically color and opacity).
Both steps are guided by domain knowledge and understanding of the visual parameters involved in rendering.
Despite years of research on the specific design and parametrization of TFs, obtaining robust TFs that can be applied to several volumes and provide consistent highlighting remains a challenge.
This is due in part to the expertise required, as both a technical understanding of the visual parameters and sufficient domain knowledge are needed to design a good TF.
The task is complicated further by variations in the data which might cause structurally different areas to hate the same intensities (or general attributes), which complicates the classification step of the TF.
To overcome this problem, much of the research in DVR has been dedicated to finding features that can discriminate structures that overlap in intensity space, introducing task-specific solutions that often do not generalize well to other use cases. Furthermore, using additional features introduces the problem of designing a multi-dimensional TF. Specification of 2D TFs already requires a lot of manual interaction and multidimensional TFs specification is a complex problem that in itself motivated many publications\cite{Ljung2016}.
One potential solution to achieve consistent colorization is using explicit semantic labels, provided by semantic segmentation or manual labels. However, training semantic segmentation models requires the annotation of typically hundreds of slices \textit{per volume} to create robust data sets. In medical applications, this has a high cost associated as labeling has to be performed by clinical experts. Furthermore, creating expressive visualizations from the raw data and labels is still not a trivial task and often requires much iteration between domain experts and visualization specialists.


The techniques we present in this work eliminate the tedious process of finding features and specifying the associated visual properties by replacing the respective steps with learned models. 
We revisit the classic DVR pipeline and systematically identify the steps that can be supported with learned deep mapping functions.
The differentiable nature of DVR allows us to train these models from 2D images, effectively reprojecting 2D annotations into 3D where meaningful features and visual mappings are derived implicitly during training. 
Our proposed process avoids the time-intensive highlighting of \emph{where} interesting structures are in 3D and instead allows experts to work directly in the rendered image space to indicate \emph{what} they want to see and \emph{how} it should appear.
This requires fewer annotations and these annotations can directly reflect the intended visual outcome.
Besides direct manual annotation, a wide range of other techniques could also be used to create the training images for our methods. For example, surface meshes could be reconstructed for a limited data set and surface shaders could be used to create a highly customized illustrative rendering. 
Our end-to-end training can incorporate the illustrative rendering aspects and apply it to unseen input data without the need for explicit surface extraction.
Our work is the first to describe neural rendering for scientific volumes while directly modeling the feature extraction and TF mapping steps within the model. 
Even though our architectures are currently constrained by relatively long training and inference times, clear benefits of our work can already be seen for the medical context.

In this paper we introduce a unified framework for direct volume rendering that enables  learning of implicit functional mappings by explicitly modeling the DVR process within the architectures.

In summary, we present the following main contributions:
\begin{itemize}
    \item A generalized formulation of DVR that allows the integration of deep neural networks into parts of the classic pipeline, thus enabling \emph{Deep Direct Volume Rendering} (DeepDVR)
    \item A set of deep volume rendering architectures which are derived from this formulation by introducing deep networks in several parts of the rendering
    \item An effective training strategy for models with explicit raymarching we call \emph{stepsize annealing}, validated by experiments
    \item Experiments to demonstrate of the effectiveness of image-based training of our deep rendering architectures on the tasks of (1) image-based TF specification and (2) learning generalized visual features in volume and image space
\end{itemize}

The paper is structured into the following sections:
Section~\ref{sec:rw} provides an overview of the related literature.
Section~\ref{sec:methods} introduces the mathematical foundation of our \emph{deep direct volume rendering} and presents various DeepDVR architecture variants. We also present DVRNet, a novel architecture for multiscale feature extraction in volume rendering tasks.
In our experiments (\autoref{sec:experiments}), we first perform a detailed analysis of DeepDVR-specific metaparameters and parametrizations for intensity transfer functions. These experiments demonstrate the effectiveness of our \emph{stepsize annealing} training method.
In \autoref{sec:deepvr}, we compare our deep architectures on two tasks: (1) learning to render from manual annotations in image space and (2) learning generalized volume rendering from a multi-volume data set.
We discuss the implications of our experiments in \autoref{sec:discussion} and provide concluding remarks in \autoref{sec:conclusion}.

                    \section{Related Work}
\label{sec:rw}

\subsection{Volume Rendering and TFs}

\cite{Max1995} has introduced the first formulation of direct volume rendering, however the GPU-accelerated pipeline described by \cite{Kruger2003} have played a crucial role in forming the pipeline for modern hardware-accelerated raymarching.
\cite{Ljung2016} provide an excellent overview of recent techniques for TF design in their state of the art report.
Multidimensional TFs, leveraging derived attributes to improve classification, have been proposed in variations. \cite{Kniss2001} have proposed the use of gradient magnitude and a directional gradient attribute in addition to intensity. 
Different derived attributes based on for example curvature \cite{Kindlmann2003}, size \cite{Correa2008a} or a local occlusion spectrum \cite{Correa2009} were also introduced. 


The complexity of two-dimensional TF specification has been often dealt with by specifying 2D geometries on the 2D feature histogram\cite{Kniss2001}. \cite{Rezk-Salama2006} have proposed the use of semantic models represented as 2D primitives in the intensity-gradient magnitude space. In their work, experts can create a basic semantic model which can then be easily adapted by non-experts to a specific data set. To make higher-dimensional spaces more tractable, \cite{DeMouraPinto2007} use Kohonen maps to reduce the high-dimensional feature space to two dimensions in which the same user interfaces can be used.
\cite{SchultezuBerge2014a} introduced a simplified approach to specify the importance of multidimensional features by using a predicate weighting for each feature. 

Machine learning has already been applied to replace the TF to some extent: \cite{Soundararajan2015} use machine learning models to learn a TF from scribbles in the volume domain. They compare five machine learning approaches including a single layer perceptron (SLP), finding random forests to be favorable due to their superior speed and robustness.
Approaches have been proposed for modifying TFs in image space via strokes to indicate areas that should be changed in the output image. \cite{Ropinski2008} use feature histograms along the rays covered by the strokes to adapt the TF. \cite{HanqiGuo2011} extended these ideas with novel interaction metaphors for contrast, color and visibility control directly in image space.

Deep learning has also been used in the context of volume rendering to synthesize novel views or views with different parameters \cite{Hong2019, He2020}, for super-resolution of volume isosurface renderings \cite{Weiss2019}, for compressed rendering of time-varying data sets \cite{Jain2017} and for prediction of ambient occlusion volumes \cite{Engel2020}. However, contrary to our work, none of these approaches explicitly take the DVR method into account directly in the model architecture.


\subsection{Differentiable Rendering}
Differentiable rendering has been a subject of recent interest in computer graphics as a building block for machine learning pipelines\cite{Kato2020}. Differentiable mesh rasterizers \cite{Liu2019,Loubet2019,Li2018,Nimier-David2019} provide interesting solutions to inverse rendering tasks, however they are generally not extendable to volumetric data.
Neural rendering\cite{Tewari2020} is a relatively novel field in which a rendering step is incorporated into the network architecture. This allows for the explicit or implicit incorporation of scene parameters into the training and enables applications like scene relighting, novel view synthesis, facial and body reenactment and photorealistic avatars, among others.

The specific combination of volume rendering with machine learning has been addressed in recent literature:
\cite{Nguyen-Phuoc2018} introduced RenderNet, a deep convnet that performs differentiable rendering of voxelized 3D shapes.
Their proposed network consists of a 3D and a 2D convolutional part connected by a novel \emph{projection unit} which combines the features along the viewing ray with an MLP. This enables the network to learn different rendering styles and was also shown to be effective to synthesize novel viewpoints for faces captured from a single viewpoint.
\cite{Rematas2020} present a controllable neural voxel renderer based on this \emph{projection unit} which produces detailed appearance of the input, handling high frequency and complex textures.
More pertinent to rendering scientific volumes, \cite{Berger2019} have formulated the volume rendering task as an image generation from a given camera and TF. They use generative adversarial networks (GANs) to effectively train a network to memorize a specific volume dataset, representing a differentiable renderer for this volume that is conditioned only on viewpoint and TF. This differentiable renderer can then be used to further explore the latent space of TFs and provide a sensitivity map visualizing areas in the output image affected by specific parts of the TF.

Raymarching-based differentiable rendering has recently gained attention in the context of inverse rendering for scene reconstruction, where different scene representations have been explored:
\cite{Lombardi2019b} uses \emph{Neural Volumes} to optimize volumetric scene representations from sets of camera images. They use an end-to-end approach to reconstruct an explicit color+opacity volume that is then rendered from the known viewpoints and compared to the reference images. 
Instead of learning an explicit volumetric representation, Scene Representation Networks \cite{Sitzmann2019} encode both geometry and appearance into the network itself as a mapping of 3D location to feature vector. The feature vector of each position encodes both the signed distance to a surface, as well as features that are later decoded to the final surface color.
\cite{Liu2020} propose an efficient differentiable sphere-tracing algorithm to render implicit signed distance functions.
\cite{Mildenhall2020a} train a network that is conditioned on a 3D position and viewing direction and returns an RGBA tuple, thus encoding a \emph{Neural Radiance Field} in the network.
This radiance field can be sampled at arbitrary points to create novel photorealistic viewpoints from a set of images while accurately representing complex geometry and materials. \cite{Niemeyer2020} follow a similar approach to learn deep implicit representations of shape and texture without 3D supervision.
Although many of these works include a volume raymarching step within the training pipeline, neural rendering has not yet been proposed for scientific volume visualization. 
        \section{Methods}\label{sec:methods}
The beneficial properties of the widely used classical emission-absorption model (o.e. \cite{Max1995}) for DVR provide a strong potential for inverse rendering tasks in scientific volume rendering.
We first recapitulate the classic formulation of DVR as a basis for our following discussions. 

\subsection{Direct Volume Rendering}
In this optical model, each position $\mathbf{x} \in \mathbb{R}^3$ in the volume is associated with an emission component $c(\mathbf{x}) \in \mathbb{R}^3$ and an absorption coefficient $\kappa(\mathbf{x}) \in \mathbb{R}$.
Computing the radiant energy $C$ incident on a virtual camera's pixel is then formulated as evaluating an integral along the camera ray $\mathbf{x}(t)$:

\begin{equation}
    C = \int_0^\infty c\left(\mathbf{x}(t) \right)\cdot e^{-\tau (t)} dt
\end{equation}

where $\tau(t') = \int_{0}^{t'} \kappa(\mathbf{x}(\hat{t}))d\hat{t}$ can be interpreted as a visibility term. 

This integral is commonly approximated numerically with ray casting, where equidistant samples at $t_i = i \Delta t$ along the ray are combined using the well-known alpha blending equation. In this case, the emitted color $C_i$ and opacity $A_i$ at sample point $i$ are defined as
\begin{align}
    C_i &= c(i \Delta t) \Delta t, \\
    A_i &= 1 - e^{-\kappa(i \Delta t)} \Delta t.
\end{align}

The step size $\Delta t$ is often chosen relative to the voxel size $d_v$:
\begin{equation} \label{eq:samplingrate}
        \Delta t = d_v / s
\end{equation} where $s$ is the \emph{sampling rate}.

Iterative evaluation can be achieved from front to back with alpha blending as
\begin{align}
\label{eq:alphablendcolor}
    C_i' &= C_{i-1}' + \left(1 - A_{i-1}'\right) C_i, \\
\label{eq:alphablendalpha}
    A_i' &= A_{i-1}' + \left(1 - A_{i-1}'\right) A_i
\end{align}
with starting conditions $C_0' = 0, A_0 = 0$.

Numerical integration of the ray with a fixed sampling rate can lead to aliasing artifacts. A common approach to reduce this is to use \emph{ray jittering} \cite{Danskin1992} which applies a random offset from a uniform distribution $t_o \sim \mathcal{U}(0,t_{j,\mathrm{max}})$ in the ray direction $\mathbf{d}$ to each pixel ray as
\begin{equation}\label{eq:jitter}
    \mathbf{x}(t) = \mathbf{x_0} + (t + t_o) \mathrm{d}
\end{equation}
where the jitter magnitude $t_{j,\mathrm{max}}$ is usually chosen as $\Delta t$.

For a more in-depth discussion of the derivations, we refer the interested reader to the course notes by \cite{Hadwiger2008a}.

\label{sec:diffdvr}
In order to perform image-based optimization of the rendering algorithm, we require an extensible differentiable mathematical model of volume rendering. We generalize the well-known emission-absorption volume rendering equations to an arbitrary number of color channels. We also introduce replaceable functions for feature extraction, color and opacity TFs, composition and color space decoding, which can be then implemented with learning-based approaches.

\subsection{Generalized Direct Volume Rendering}
\begin{figure}
    \centering
    \includegraphics[width=\columnwidth]{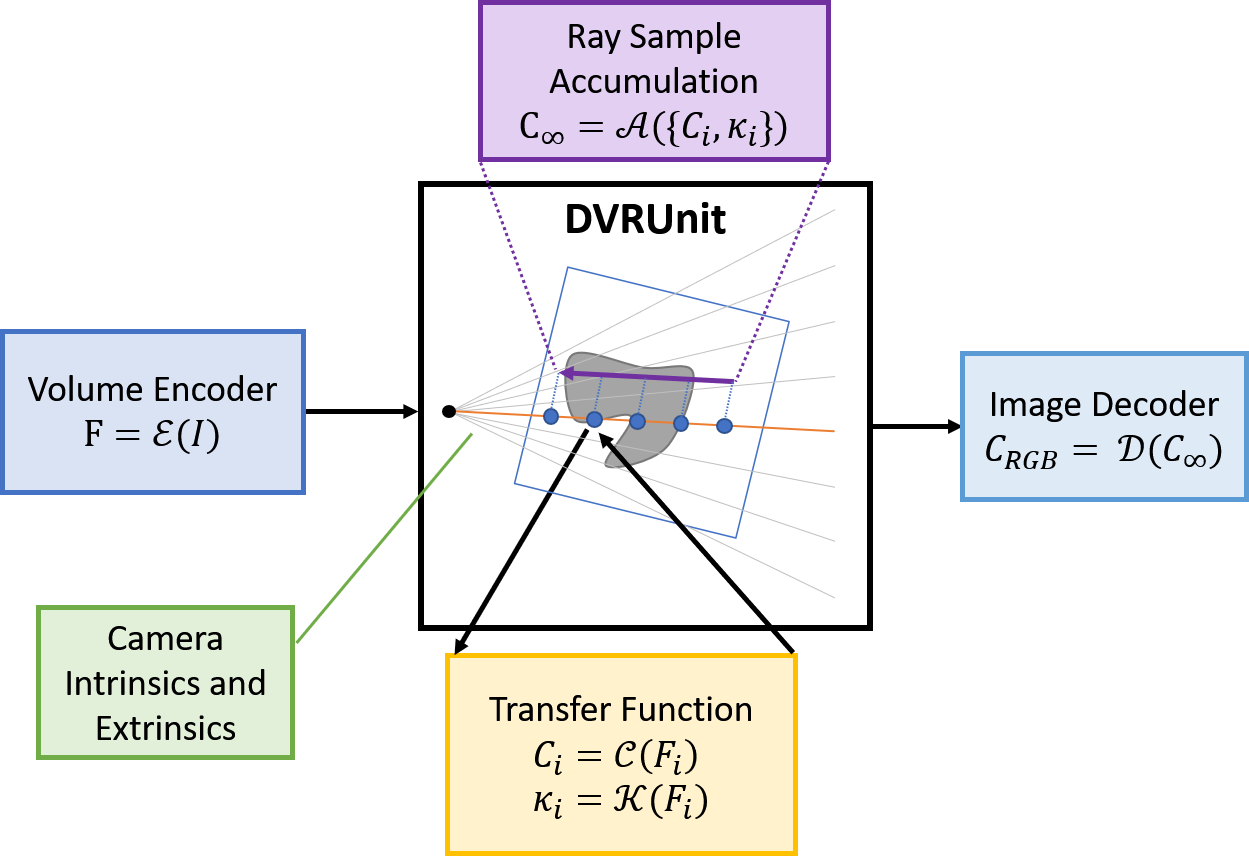}
    \caption{Schematic overview of the proposed DVR Unit based on our generalized DVR formulation. Our architectures derive from this by implementing these functional mappings with deep neural networks. }
    \label{fig:DeepDVR}
\end{figure}

Our novel deep volume rendering method is based on two observations: Firstly, the volume rendering method is fully differentiable, making it possible to integrate it with deep learning architectures using automatic differentiation frameworks \cite{Paszke2017}.
Secondly, these formulations are independent of the color space in which the computations are performed. Commonly, the emissive component $c$ is represented as a linear RGB triplet, however it is possible to evaluate these in any linear color space.
We formulate our deep ray casting algorithm using a generic input feature space $\bm{F} \in \mathbb{R}^{n_F}$ and an $n_c$-dimensional color space $c \in \mathbb{R}^{n_c}$.
Our deep volume rendering pipeline consists of three distinct parts: (1) feature extraction, (2) deep volume rendering by volume sampling and subsequent sample accumulation and (3) image decoding (c.f. \autoref{fig:DeepDVR}).

The purpose of input feature extraction is to derive further features to discriminate between different classes in order to assign different optical properties.
We formalize this by introducing a \emph{volume encoder} function $\mathcal{E}(I): \mathbb{R}^{n_I,H,W,D} \rightarrow \mathbb{R}^{n_F,H,W,D}$ that transforms the scalar input volume $I(\mathbf{x})$ to an $n_F$-dimensional feature space:
\begin{equation}
    F(x) = \mathcal{E}(I(\mathbf{x}))
\end{equation}

We split the TF into two parts and introduce $\mathcal{C}(F): \mathbb{R}^{n_F} \rightarrow \mathbb{R}^{n_C}$ and $\mathcal{K}(F): \mathbb{R}^{n_F} \rightarrow \mathbb{R}$ which compute the emission $c \in \mathbb{R}^{n_C}$ and absorption coefficient $\kappa$, respectively, from the feature space. This yields the emission $C_i$ and absorption $A_i$ for each sampling point:
\begin{align}\label{eq:deeptf}
    F_i &= F(\mathbf{x}(i \Delta t)),            \\
    C_i &= \mathcal{C}(F_i)\Delta t,               \\
    A_i &= 1 - e^{-\mathcal{K}(F_i)}\Delta t.
\end{align}

The samples $(C_i, A_i)$ of a ray are then combined with a \emph{accumulation function} $\mathcal{A}$ to a final ray color:
\begin{equation}\label{eq:deepcompositing}
    C'_{\infty} = \mathcal{A}\left(\left\{ (C_i, A_i) | i \right\}\right).
\end{equation}
Within this arbitrary $n_C$-dimensional color space, alpha blending equations can still be evaluated as described per equations \ref{eq:alphablendcolor}-\ref{eq:alphablendalpha}. However, $\mathcal{A}$ could also represent illustrative or importance-based compositing operations \cite{Bruckner2009,DeMouraPinto2010a}.

The resulting image then consists of the $n_C$-dimensional projected features $C_\infty$ and an alpha channel $A_\infty$. We therefore introduce an \emph{image decoder} function $\mathcal{D}(F)$ to transform the rendered features back into an RGB color space via
\begin{equation}
    C_\infty^{RGB} = \mathcal{D}(C_\infty').
\end{equation}


            \subsection{TFs based on MLPs}
\label{sec:mlptf}
In scientific volume visualization, TFs are routinely used to map from the voxel feature domain $F(x)$ to the optical properties (emissive color $c$ and absorption coefficient $\kappa$).
This mapping strongly influences the visual aspects of the rendering (c.f. \autoref{fig:classic_tf_workflow}), controlling both which structures are relevant (\textit{classification}) and how these structures appear visually (\textit{visual mapping}).
As such, the TF is usually parametrized such that users can modify and tweak it to a specific data set in order to achieve the desired effect.
In practice, TFs are often represented as lookup tables or through simple primitives like trapezoid and parabolic functions (\cite{Rezk-Salama2006}) or a sum of Gaussian functions (\cite{Kniss2003}).
Manual specification of traditional TFs is already tedious in one- and two-dimensional feature spaces and has to rely on dimensionality reduction techniques for higher dimensions to make the problem even tractable from a user interaction perspective\cite{DeMouraPinto2007}.
Yet, TFs mathematically are merely a mapping from an $n_F$-dimensional input to a 4-dimensional (RGB + $\kappa$) output space for the emission+absorption DVR model and can therefore be approximated with multi-layer perceptrons (MLP). 



By replacing the traditional TF representation with an MLP, we eliminate the need for directly modifying the mapping function manually, as the MLP parameters for the most part do not allow for meaningful direct manual manipulation.
Instead, these parameters are optimized mathematically given example data. Given the differentiability of the DVR in the formulation outlined above, these parameters are optimized as part of the proposed end-to-end training procedure. This allows us to provide training data in image space and rely on gradient descent to optimize the MLP weights instead.

    \subsection{Deep Direct Volume Rendering}
\label{sec:deepvr}
The generalized reformulation of the direct volume rendering algorithm above was also motivated by the goal to eliminate the explicit design of application-specific features in a general way by using a convolutional neural network (CNN) for the feature extractor $\mathcal{E}$. Our algorithm also allows for creating architectures that perform raymarching in a latent higher-dimensional color space and then transform this high-dimensional image to the RGB color space.


\label{sec:architectures}
In the following, we outline a set of architectures derived from our general formulation of direct volume rendering in Section~\ref{sec:diffdvr}.
We have designed several architectures based on replacing the generalized functions for feature extraction $\mathcal{E}(I)$, color and opacity TFs $\mathcal{C}(F), \mathcal{K}(F)$ and the image decoder $\mathcal{D}(C)$.

\begin{figure*}
    \centering
    \includegraphics[width=\textwidth]{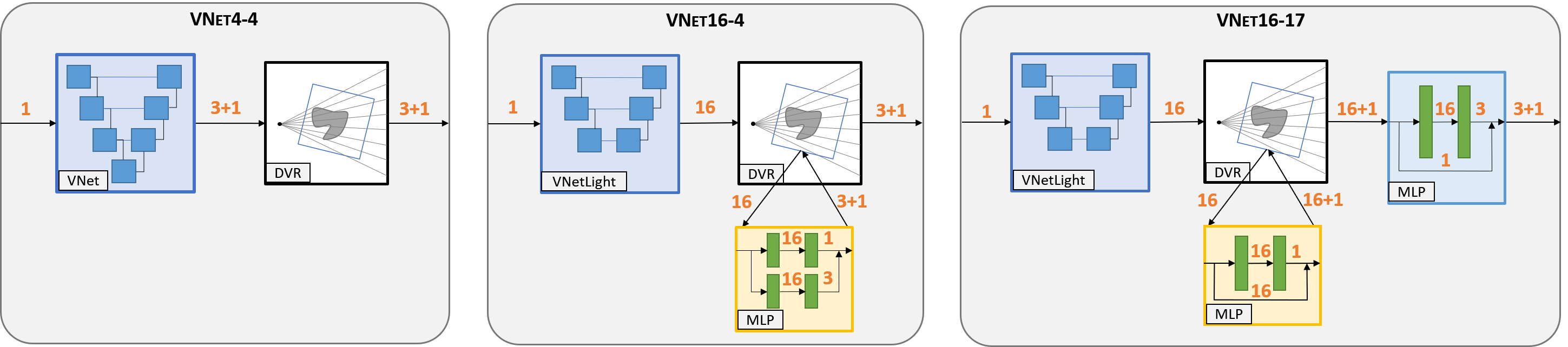}
    
    
    \caption{
        Our three VNet-based model architectures with increasing complexity. \textsc{VNet4-4} uses a deep encoder network to directly infer the $(RGB,\kappa)$ volume to be rendered. \textsc{VNetL16-4} creates 16 semantic channels in the encoder which are mapped to color and opacity via an MLP-based TF. \textsc{VNetL16-17} instead uses an MLP to map these channels to 16 color space channels, which are then decoded with a final MLP on the 16-channel output image.
        "+1" indicates a separately handled opacity/alpha channel.
    }
    \label{fig:architectures}
\end{figure*}

Our architectures, summarized in Table~\ref{tab:archsummary} and Fig.~\ref{fig:architectures}, are progressing in complexity by replacing more of the functions with deep neural networks. For comparison, we also include RenderNet~\cite{Nguyen-Phuoc2018} as a baseline in our experiments as, among previously published works, this architecture is most closely related to our concepts.
RenderNet resamples the input volume to view space using the perspective camera transform, such that the X-axis of the resampled volume corresponds to the ray directions and the YZ-axes correspond to the image coordinates. This resampled volume is then processed by a 3D convolutional network (CNN) to extract semantic features, reducing the spatial dimension in the process. This step corresponds to our \emph{input encoder} and \emph{opacity and color TFs}. RenderNet uses an MLP to project all voxels along the X-axis to a single high-dimensional feature vector. This \emph{projection layer} is analogous to our generalized \emph{accumulation function} $\mathcal{A}$. The resulting 512-channel 2D projection is then upsampled to the target resolution using a 2D CNN, similar to our \emph{image decoder} function $\mathcal{D}(C)$.
There are some key differences between RenderNet and our approach of deep direct volume rendering: Firstly, the view space resampling in RenderNet is performed at the beginning of the pipeline whereas in our approach, the volume encoder extracts features in object space, making them independent of the camera parameters. Secondly, our formulation uses a more explicit modeling of occlusion through the absorption coefficient and alpha blending. Thirdly, we avoid excessive downsampling in the spatial dimensions by using specific architectures that employ skip connections.

We introduce four novel architectures based on these considerations: three VNet-based architectures (\textsc{VNet-4-4}, \textsc{VNetL-16-4}, \textsc{VNetL-16-17}) and  \textsc{DVRNet}, a novel multiscale rendering architecture.

 \begin{table*}
 \ra{1.1}
   \caption{
        Summary of novel architectures and comparison to baseline methods.
         $\mathcal{I}_N(x) = x$ is an N-dimensional identity function.
        "Alpha" designates alpha blending. \textsuperscript{*} DVRNet uses DVR at multiple scales and does not fit directly into this categorization.
    }
   \label{tab:archsummary}
   \begin{minipage}{\textwidth}
   \begin{center}
   \begin{tabular}{lcccccr}
     \toprule
                    & $\mathcal{E}(I)$  &  $\mathcal{C}(F)$ & $\mathcal{K}(F)$  & $\mathcal{A}(\{(C_i, A_i\})$ & $\mathcal{D}(C)$ & Parameters\\
     \midrule
    Lookup TF       & $\mathcal{I}_1$          & Lookup               & Lookup           & Alpha         & $\mathcal{I}_3$    &  \textasciitilde 1 K\\
    RenderNet       &  \multicolumn{3}{c}{\textasciitilde \textasciitilde \textasciitilde \textasciitilde \textasciitilde \textasciitilde 3D CNN \textasciitilde \textasciitilde \textasciitilde \textasciitilde \textasciitilde \textasciitilde}                                        & $\mathrm{MLP}_{512}$ & 2D CNN      &  \textasciitilde 226 M\\ 
    \textsc{VNet4-4}        & $\mathrm{VNet}_4$        & $\mathcal{I}_3$      & $\mathcal{I}_1$  & Alpha         & $\mathcal{I}_3$    &  \textasciitilde 45.6 M\\
    \textsc{VNetL16-4}       & $\mathrm{VNet}_{16}$    & $\mathrm{MLP}_3$     & $\mathrm{MLP}_1$ & Alpha         & $\mathcal{I}_3$    &  \textasciitilde 12.3 M\\ 
    \textsc{VNetL16-17}      & $\mathrm{VNet}_{16}$    & $\mathcal{I}_{16}$   & $\mathrm{MLP}_1$ & Alpha         & $\mathrm{MLP}_{3}$ &  \textasciitilde 12.3 M\\ 
    \textsc{DVRNet}\textsuperscript{*} & VNet Blocks    & $\mathcal{I}$        & $\mathrm{MLP}$   & Alpha         & UNet Blocks        &  \textasciitilde 25.0 M \\ 
   \bottomrule
 \end{tabular}
 \end{center}
\end{minipage}
\end{table*}

\begin{figure}
    \centering
    \includegraphics[width=\columnwidth]{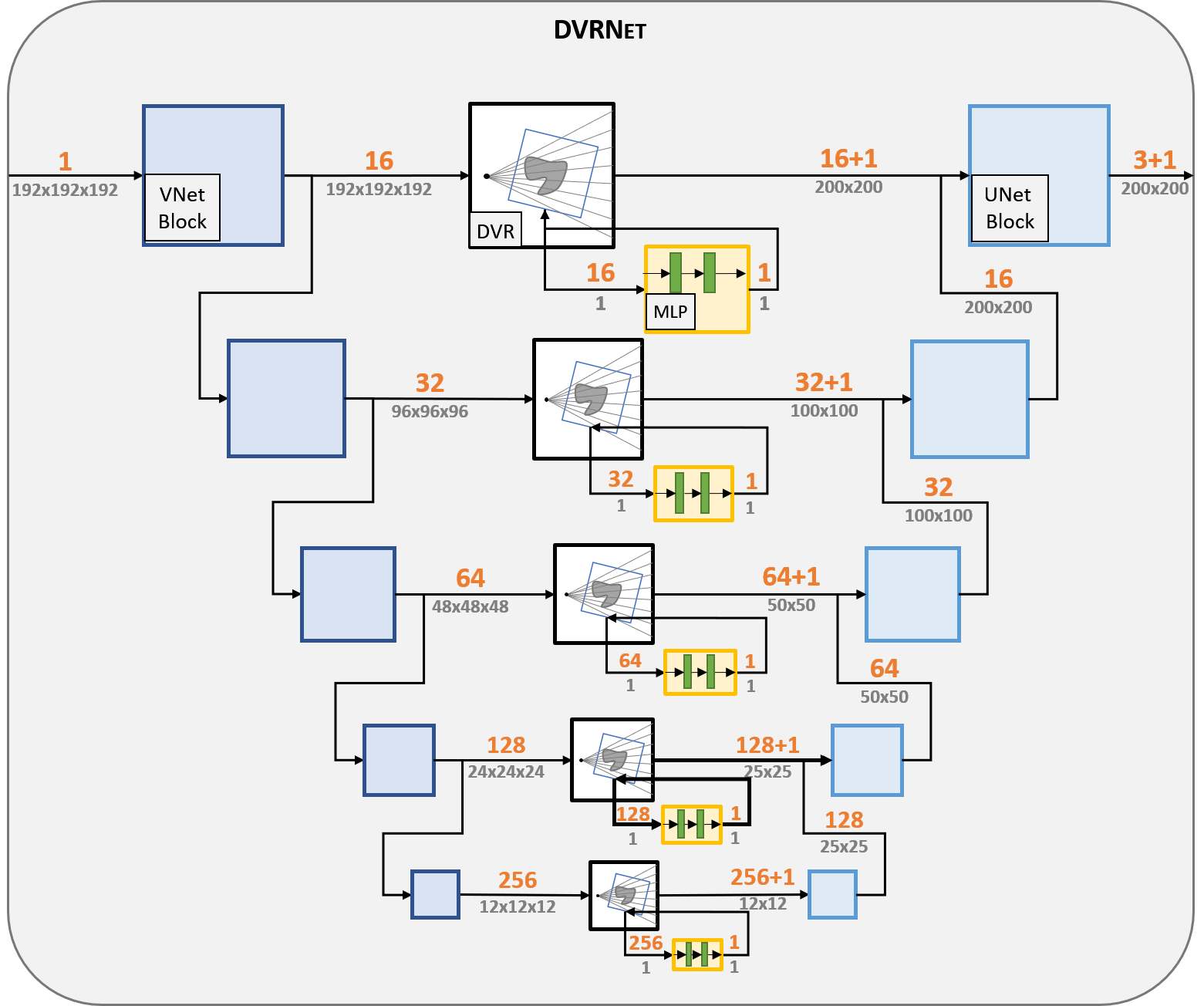}
    \caption{
        Schematic overview of our novel \textsc{DVRNet} architecture. Inspired by UNet and VNet architectures, \textsc{DVRNet} features a multiscale volumetric encoding part and a multiscale 2D decoder. Corresponding encoder-decoder levels are connected with a DeepDVR module, replacing the skip connections of the classic UNet/VNet architectures. Indicated dimensions are for the training data set, the convolutional and DeepDVR layers support arbitrary input and output dimensions during and after training.
        "+1" indicates a separately handled opacity/alpha channel.
    }
    \label{fig:dvrnet}
\end{figure}

\paragraph{\textsc{VNet4-4}} The first architecture was created by implementing the \emph{volume encoder} $\mathcal{E}(I)$ with a 3D convolutional deep neural network.
We chose to use an existing, well established multiscale encoder-decoder architecture called VNet\cite{Milletari2016} which has been shown to perform well for binary volumetric image segmentation. 
We adapt this network to a four channel output to obtain an RGB$\kappa$ volume directly.
This essentially implements a \emph{pre-classified} volume rendering, for which \cite{Engel2001} have discussed further rendering optimizations we do not apply here, but which could be used outside of training.

\paragraph{\textsc{VNetL16-4}} This architecture is extended from the previous by replacing the VNet network with a simplified "VNetLight" variant with one less level of downsampling to reduce the number of parameters, and by producing 16 semantic channels instead.
Additionally, we add MLP-based trainable color and opacity TFs. For each of the two TFs, an MLP is defined that feeds the 16-channel input to a 16-channel hidden-layer with a ReLU activation and then produces either 3 channels (RGB) or a single channel ($\kappa$) using a sigmoid activation function. Again, no \emph{image decoder} is required as the image is produced in the RGB color space.

\paragraph{\textsc{VNetL16-17}} This architecture performs compositing in an abstract 16-channel color space. As before, a VNetLight architecture implements the \emph{volume encoder} to produce a 16-channel volume. For $\mathcal{K}(F)$, the same MLP architecture as with VNetL16-4 is used. Instead of a color TF, the 16 semantic channels are directly used in compositing (i.e. $\mathcal{C}(F) = F$). To decode the resulting 16-channel image (with one additional alpha channel), we use another MLP with one ReLU-activated 16-channel hidden layer and a sigmoid function in the output layer, yielding the 3-channel RGB image whereas the opacity is used directly as produced from the DVRUnit.

\paragraph{\textsc{DVRNet}} Our architecture is inspired by the encoder-decoder architectures that have shown great success in image segmentation due to their ability to use both local and global semantics by their multiscale nature.
Our architecture, visualized in Fig.~\ref{fig:dvrnet}, is based on the idea that the volumetric multiscale encoder part of a VNet can be connected with the upsampling decoder part of a U-Net by inserting a DVRUnit into the skip connections to perform a projection from 3D to 2D feature space.
Each level viewed on its own is a volume raymarching network similar to \textsc{VNetL16-17}, rendering a potentially spatially downsampled volume in a latent color space at a reduced image resolution.
These lower-resolution multichannel images are then combined in the U-Net decoder blocks to produce the final 2D image at the desired resolution.

This architecture was designed to achieve a similar multiscale classification power as using VNet directly as an encoder, without the need to perform full volumetric upsampling, performing the upsampling step in image space instead.

\subsection{Stepsize Annealing}
The sampling rate $s$ has a major influence on the computational footprint of the algorithm: it linearly scales the number of samples classified and composited for all rays. 
When optimizing weights with automatic differentiation, this scaling also applies to the backpropagation step and memory consumption.
Given that increasing the sampling rate severely increases training time at diminishing returns, we propose a novel training strategy for a more efficient training of differentiable volume rendering models: 
In order to benefit from the better convergence efficiency at low sampling rates, we propose a strategy that varies the sampling rate in a range $[s_l, s_h]$ across the epochs during training relative to the training progression where $e$ is the current epoch and $E$ the total number of epochs:
\begin{equation}
    s(e) = s_{l} (1 - (e/E)^2) + (e/E)^2 s_{h}. \label{eq:srprogression}
\end{equation}

This progressive supersampling of the volume during training results in a fast convergence with low sampling rates in first epochs for a rough approximation and increases the sampling rate towards the end to refine the optimized TF.

\section{Experiments}\label{sec:experiments}
Our differentiable volume rendering enables the end-to-end optimization of rendering parameters, yielding models that do not require tedious manual manipulation of transfer function parameters.

In the following sections, we evaluate the effectiveness of the presented concepts.
We first present experiments regarding the direct learning of the functional mapping $F \rightarrow (C,\kappa)$ in image space, comparing the performance of TF representations. 
In these experiments, we also evaluate how the hyper-parameters specific to volume rendering affect training and testing performance.
In the second set of experiments, we compare the our deep architecture variants for volume rendering by training on manually adapted reference images for a single volume (\autoref{sec:ex_handpainted}) and generalizing across multiple volumes to create a renderer that is robust against inter-volume variations (\autoref{sec:ex_generalization}).

    \subsection{Image-Based TF Optimization}
\label{sec:ibtfo}

In this section, we evaluate the benefits of optimizing 1D lookup TFs and MLP-based TFs in a task of image-based TF definition.


As a straightforward benchmark for our methods, we reconstruct the TF from a set of images for a single volume such that the images produced  by an optimized TF are as close as possible to the reference images. While a task like this does not directly have practical applications as-is, it serves here as an evaluation framework with well-defined, perfect ground truth to investigate the training behavior of image-based TF optimization with respect to several parameters.

\subsubsection{Experimental setup}
We train two different TF representations: a classic lookup TF where the 256 individual elements are optimized, and an MLP representation with two hidden layers. Both representations have a similar number of trainable parameters and we have confirmed that the MLP is deep enough to represent the TFs used in this experiment.
The goal of this experiment is to analyze the influence of hyperparameters that are specific to DeepDVR in order to optimize the training process. We evaluate how training with different sampling rates affects training performance and validate our novel \emph{stepsize annealing} scheme. For the Lookup TF, we further analyze on whether ray jitter (c.f. \autoref{eq:jitter}) has an effect on training.
In summary, our experiment conditions are: \emph{model} (Lookup, MLP), \emph{sampling rate} (fixed $s \in \{0.25, 0.5, 1.0, 2.0, 3.0\}$ vs. stepsize annealing with $s_l = 0.1, s_h = 2.0$) and \emph{ray jitter} (yes/no).

\begin{figure}
    \centering
    \includegraphics[width=\columnwidth,trim=0cm 0cm 3.5cm 2cm,clip]{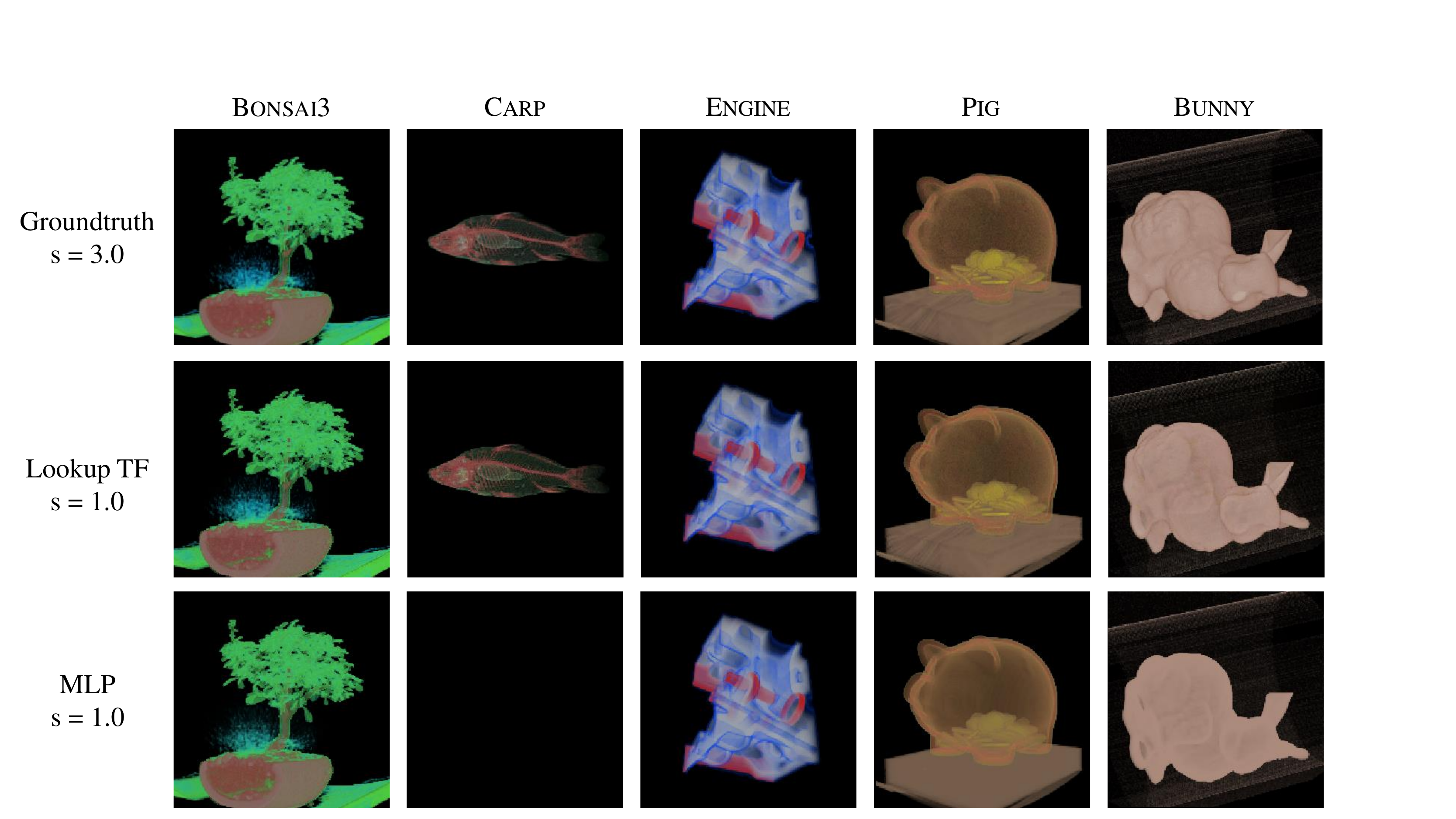}
    \caption{The five volume data sets rendered with manually adapted reference TF (top row), reconstructed lookup TF (middle row) and deep TF represented as an MLP. $s$: sampling rate. }
    \label{fig:tflearn_datasets}
\end{figure}

\paragraph{Data sets}
We create five training data sets from five different volumes sourced from the volume library\cite{Rottger2020} (c.f. \autoref{fig:tflearn_datasets}, top row) for each of which we manually designed a 1D TF. 
Each of the volumes was converted to floating point, resampled and padded to an isotropic voxel resolution of $256 \times 256 \times 256$, and then a training set of 25 images was created with a normal DVR renderer using the manually defined TF, with an additional 7 views as a validation set.

\paragraph{Metrics}
We use structural similarity (SSIM)\cite{Wang2004} to assess image quality after every epoch.
To evaluate the perceptual quality of the produced images, we report two perceptual similarity metrics: The \textit{Fréchet Inception Distance} (FID) score \cite{Heusel2017} and \textit{Local patch-wise image similarity} (LPIPS)\cite{Zhang2017}. Both metrics are based on auxiliary, pre-trained neural networks to evaluate perceptual similarity instead of per-pixel comparisons.
The specified sampling rate $s$ in the charts is exclusively used for training. During testing, we render all images using $s=3.0$ to provide a fair comparison with the ground truth. 
We also measure the total time required for training all epochs on our hardware as an indicator for the computational scaling.

\begin{figure}
    \centering
    \includegraphics[width=\columnwidth]{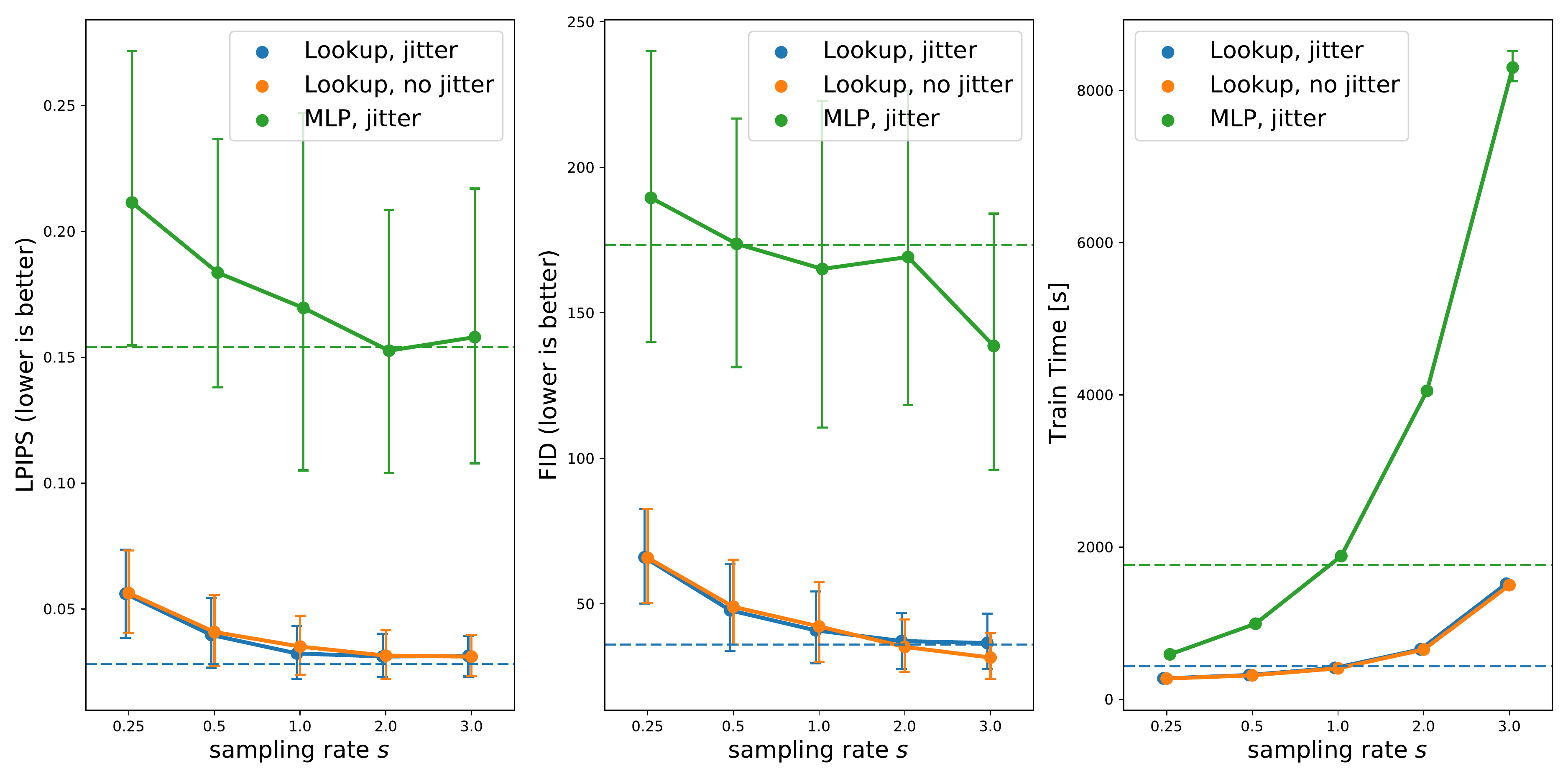}
    \caption{Training behavior with respect to sampling rate. Graphs show average metrics and 95\% CI, horizontal lines indicate results for our proposed \emph{stepsize annealing} training strategy. Each data point summarizes 25 trainings on the 5 data sets.}
    \label{fig:tflearn_analysis}
\end{figure}

\paragraph{Training}
We use the Adam~\cite{Kingma2015} optimizer ($\beta_1=0.9, \beta_2 = 0.99$) at a learning rate of 0.3 for 100 epochs. SSIM is tracked for every epoch and we retain the model with the highest value.
We performed MLP training with the same set of training conditions, however only train with ray jittering enabled. 
We use a fixed learning rate of 0.3 for Lookup and 0.05 for MLP models in all runs. The batch size was chosen manually for each of the sampling rates as higher sampling rates require more GPU memory. We used batch sizes of $\{12, 6, 3, 2, 1\}$ for the sampling rates $\{0.25, 0.5, 1.0, 2.0, 3.0\}$ respectively.
Our models are implemented with with Pytorch. All trainings were performed on an NVidia Tesla V100 32GB, 2x Intel Xeon Gold 5120 (2x14 cores) and 256 GB of system RAM.

We refer the reader to our supplementary document, where we provide more details on the training of these experiments as well as more detailed results.

\subsubsection{Results}
For each of the five datasets, we repeat training five times for each condition with random initialization and report the average across these 25 runs in \autoref{fig:tflearn_analysis}. The best of five runs at training $s=1.0$ can be observed in \autoref{fig:tflearn_datasets}.

From \autoref{fig:tflearn_analysis}, it can be seen that training time increases with the sampling rate, however the perceptual similarity measures show diminishing returns when increasing the sampling rate beyond 1.0.
This is the main motivating argument for our \emph{stepsize annealing} strategy, which is shown as horizontal lines in the plots. For both models, \emph{stepsize annealing} produces visual results comparable to training with a fixed sampling rate of $s=2.0$ however at an average reduction of 33.4\% in training time compared to a constant sampling rate of $s=2.0$.

Comparing the results with and without jittering there is a slight improvement for low sampling rates (0.25, 0.5) when using jittering. However, the benefit of ray jitter vanishes with higher sampling rates.
We assume that ray jittering improves the results for low sampling rates because jittering causes the same ray to cover different locations in 3D space which are otherwise skipped over at low sampling rates. Thus, using ray jittering when undersampling the volume increases the range of intensity samples taken into account across the epochs. 

From our results it is evident that the MLP parametrization performs considerably worse than the trained Lookup table. Average training time is increased by a factor of 2.0 (for $s=0.25$) to 6.19 (for $s=2.0$) due to the more expensive evaluation of the MLP layers. The visual metrics show a very high variability which we attribute to the random initialization of the MLP weights, showing a high dependence on the initialization.

The qualitative results in \autoref{fig:tflearn_datasets} (bottom row) show that in principle the MLP produces similar rendering for most data sets. Especially the \textsc{Bonsai3} and \textsc{Engine} data sets perform well. 
\textsc{Pig} and \textsc{Bunny} exhibit slightly stronger differences and \textsc{Carp} is a clear failure case. 

The \textsc{Carp} data set results highlight a general limitation of image-based TF optimization: once the TF reaches a fully transparent state, it cannot recover from it. Closer inspection of the intermediate results during training revealed that the MLP converges to a black image after 5-6 epochs, with a constant loss in the remaining epochs. Due to the use of MSE as a loss function and the very dark and highly transparent target images, a black image caused by a fully transparent TF is a strong local minimum of the loss function.
However, once the resulting images are fully transparent, gradient back-propagation is unable to recover to a non-transparent image: because no samples along a ray are contributing to the final color, no gradient update will be propagated through the samples into the TF and thus the optimization will be stuck in the local minimum.
Lookup TFs in general are susceptible to the same problem but there is less incentive to subdue all opacity parameters $\kappa_i$ to zero at the same time due to the more targeted way parameters affect the intensity mapping in this parametrization.

Overall, our results suggest that replacing TFs with equivalent MLPs is not effective for scalar-valued feature spaces.
Nonetheless, MLP-based TFs can still be feasible for higher-dimensional feature spaces where both manual specification and storage of lookup-table-based TFs become impractical.

                \subsection{Handpainted Reference Image Inversion}\label{sec:ex_handpainted}
To demonstrate the power of deeper architectures to learn complex classification features, we define a task that involves learning from user-adapted images in a workflow as shown in \autoref{fig:teaser_pipeline}.
This is similar to the scribble-based TF interactions that were proposed in previous work \cite{HanqiGuo2011,Ropinski2008}, where users can modify the TF via interactions in the 2D image space.

We have created two new datasets based on the original \textsc{Bonsai3} and \textsc{Pig} data sets consisting of 32 images of $200 \times 200$ resolution.
The 32 reference images were rendered with an initial intensity-based TF and subsequently edited in an image editor to create an exemplary semantic colorization: 
For creating the \textsc{Bonsai-H} data set, the spurious green structures from the CT scanning table were removed, the color of the flower pot and the tree's stem were adapted to consistent brown and orange tones and the small grass at the base of the tree was colored blue. In the \textsc{Pig-H} data set, the visibility and brightness of the coins was adapted to a bright yellow and the coin slot at the top of the piggy bank was colored green.
Manually designing a traditional (multidimensional) TF to achieve these visual effects would be extremely difficult to achieve even with well-designed application-specific input features.


\subsubsection{Experiment setup}
We optimize a rendering model for one specific volume and as such, the resulting rendering model does not necessarily generalize to other data sets in this case. 

In this task, we used a combination of MSE and SSIM as our loss function as we found the SSIM in the loss function helpful to recreate finer details:
\begin{equation}
    \mathcal{L}(Y, Y_{\mathrm{pred}}) = \mathcal{L}_{\mse}(Y, Y_{\mathrm{pred}}) + (1 - SSIM(Y, Y_{\mathrm{pred}})).
\end{equation}


We use the same data split as in Sec.~\ref{sec:ibtfo} and train for 200 epochs. 
We adapted learning rate and batch size for the models in preliminary experiments, maximizing the batch size with respect to the available 32GB of GPU memory for each model. Lookup (bs=32, lr=0.03) and RenderNet ($bs=10$, $lr=0.05$) can train with higher batch sizes and learning rates while all VNet* variants and \textsc{DVRNet} are restricted to $bs=2$ and a $lr=0.003$.

\subsubsection{Results}

\begin{figure*}
    \centering
        \begin{tabular}{>{\centering\bfseries}m{0.12\textwidth} *{5}{>{\centering}m{0.12\textwidth}} >{\centering\arraybackslash}m{0.12\textwidth}}
                \includegraphics[width=0.13\textwidth]{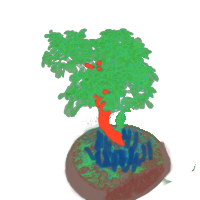} &
                \includegraphics[width=0.13\textwidth]{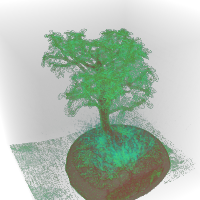} &
                \includegraphics[width=0.13\textwidth]{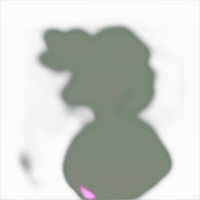} &
                \includegraphics[width=0.13\textwidth]{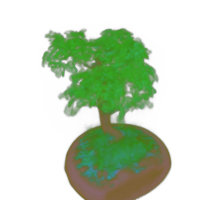} &
                \includegraphics[width=0.13\textwidth]{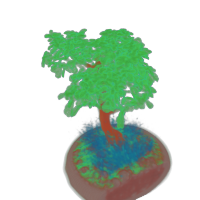} &
                \includegraphics[width=0.13\textwidth]{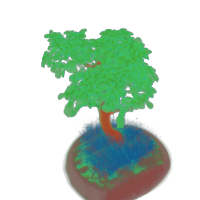} &
                \includegraphics[width=0.13\textwidth]{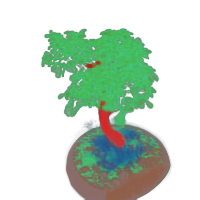} 
            \\
                \includegraphics[width=0.13\textwidth]{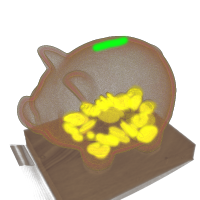} &
                \includegraphics[width=0.13\textwidth]{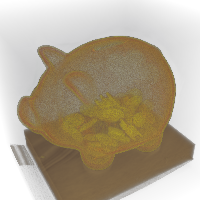} &
                \includegraphics[width=0.13\textwidth]{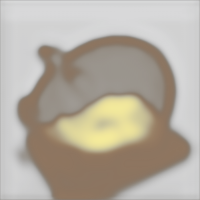} &
                \includegraphics[width=0.13\textwidth]{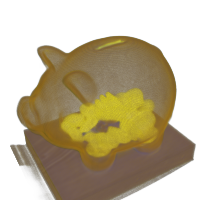} &
                \includegraphics[width=0.13\textwidth]{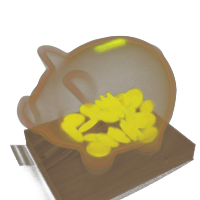} &
                \includegraphics[width=0.13\textwidth]{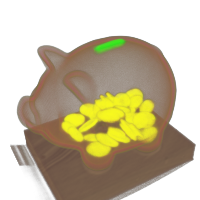} &
                \includegraphics[width=0.13\textwidth]{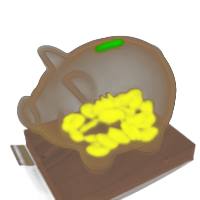}  \\
               Groundtruth &
                Lookup &
                RenderNet &
                \textsc{VNet4-4} &
                \textsc{VNetL16-4} &
                \textsc{VNetL16-17} &
                \textsc{DVRNet}
             \\
        \end{tabular}
    \caption{
        Results of our deep models trained on our manually adapted training images \textsc{Bonsai3-H} (top) and \textsc{Pig-H} (bottom). Results show a single view of the validation set, more examples can be found in the supplementary document.  Images best viewed in color at high resolution.
    }
    \label{fig:results-manual-visual}
\end{figure*}

The qualitative results in Fig.~\ref{fig:results-manual-visual} clearly show that optimizing the lookup table representation cannot find an adequate TF, instead converging to minimum which achieves only the most general desired visual qualities.
The RenderNet architecture evidently is also not suitable for this task. The excessive overblurring in image space is likely caused by the low image resolution in which the projection is performed, where spatial structures in the image can only roughly be reconstructed from the feature vector. Also, the lack of explicit opacity handling requires the model to learn a similar operation in the MLP projection layer, a task that is seemingly too complex to learn even for the with the higher number of learnable parameters of this model. This leads to inconsistent coloring between different views.
\textsc{VNet4-4} manages to assign the general semantics of the green leaves, the flower pot and the yellow coins. 
It however missed the different coloring of the tree stem and the grass, and cannot reproduce the brightness of the coins inside the piggy bank.
\textsc{VNetL16-4} and \textsc{VNetL16-17} produce very similar results on the \textsc{Bonsai3-H} data set and only differ on the \textsc{Pig-H} results. Both models successfully manage to emphasize the region of the coin slot, however only \textsc{VNetL16-17} also correctly assigns green color to it. 
\textsc{DVRNet} performs well on both data sets and reproduces the intended visual attributes in both cases. It picks up better on the distinct coloring of the tree stem, yet has problems reproducing a fully consistent green coloring from all views in the case of the coin slot.

The rough ordering from the qualitative results is reflected in the quantitative results summarized in \autoref{tab:results-manual}.
\textsc{DVRNet} and \textsc{VNetL16-17} perform very similar regarding LPIPS and FID metrics whereas \textsc{VNetL16-4} performs best in terms of SSIM on the \textsc{Bonsai3-H} data set. 

In terms of training time, \textsc{DVRNet} shows considerably better speed compared to the full VNet* variants, which can be attributed to the reduced cost of performing the upsampling in image space instead of in volume space.
The results of this experiment show that the incremental additions to the models represented by the increasing complexity of \textsc{VNet4-4, VNetL16-4, VNetL16-17} and \textsc{DVRNet}, successfully improve the perceptual capabilities of the rendering models. Overall, \textsc{DVRNet} seems to perform best in this task when also considering the training performance.

\begin{table}
\ra{1.0}
\caption{
    Results training our deep architectures on the manually adapted data sets \textsc{Bonsai3-H} and \textsc{Pig-H}.
    \textsuperscript{1}SSIM was used as part of the loss function.
}
\label{tab:results-manual}
    \begin{minipage}{\columnwidth}
    
    \begin{center}
    \begin{tabular}{@{}rc rrrr @{}}
    \toprule
    
    &&  \multicolumn{4}{c}{\textsc{Bonsai3-H}} \\
    \cmidrule{3-6} 
    && LPIPS $\downarrow$ & FID $\downarrow$ & SSIM $\uparrow$\textsuperscript{1} & Time \\
    \midrule
    Lookup &&        0.2919 & 234.47 & 0.530 &      6m \\                   
    RenderNet &&     0.4943 & 274.48 & 0.268 &  1h 33m \\                   
    \textsc{VNet4-4} &&       0.1611 & 208.04 & 0.828 &  8h 32m \\                   
    \textsc{VNetL16-4} &&     0.1001 & 148.80 & \textbf{0.923} & 9h 45m \\           
    \textsc{VNetL16-17} &&    0.1031 & 170.67 & 0.921 &  9h 46m \\                   
    \textsc{DVRNet} &&        \textbf{0.0800} & \textbf{136.52} & 0.913 & 2h 25m \\  

    \midrule
    
    &&  \multicolumn{4}{c}{\textsc{Pig-H}} \\
    \cmidrule{3-6} 
    &&    LPIPS $\downarrow$ & FID $\downarrow$ & SSIM $\uparrow$\textsuperscript{1} & Time \\
    \midrule
    Lookup &&        0.2481 & 163.74 & 0.637 &       6m \\ 
    RenderNet &&     0.3368 & 274.04 & 0.430 &  1h 33m  \\ 
    \textsc{VNet4-4} &&       0.1940 & 156.95 & 0.872 & 8h  8m  \\  
    \textsc{VNetL16-4} &&     0.1247 & 112.20 & 0.928 & 9h 54m  \\  
    \textsc{VNetL16-17} &&    \textbf{0.0830} & 106.51 & \textbf{0.932} & 9h 53m  \\ 
    \textsc{DVRNet} &&        0.0990 & \textbf{93.36} & 0.927 & 2h 36m  \\ 
    
    \bottomrule
    
    \end{tabular}
    \end{center}

    \end{minipage}

\end{table}

          \subsection{Generalized Rendering Models}\label{sec:ex_generalization}

\begin{figure}
    \centering
    \hfill
    \includegraphics[width=0.49\columnwidth]{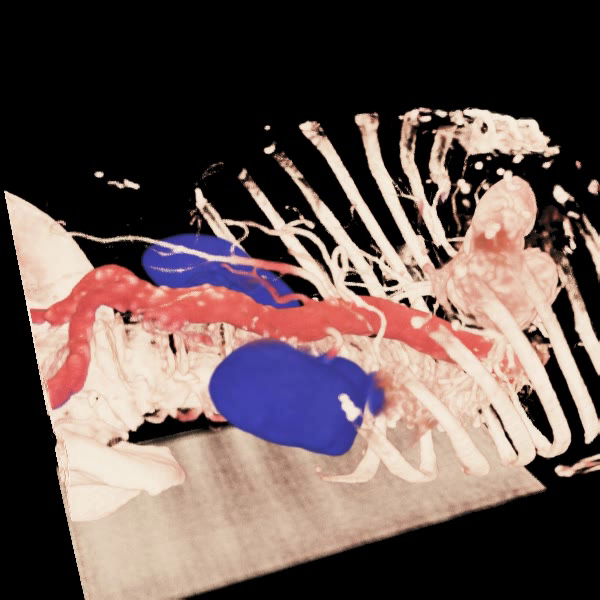} 
    \hfill
    \includegraphics[width=0.49\columnwidth]{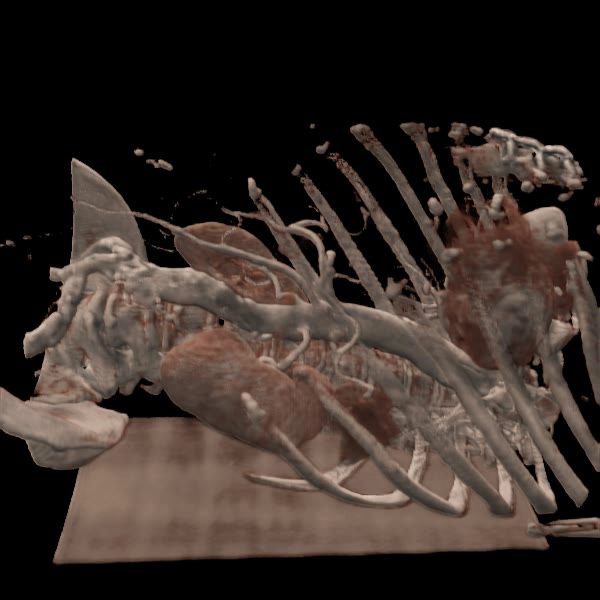}
    \hfill
    \caption{
    Our models can generate high resolution output images even though being trained only on low resolution data training data. Left:  \textsc{VNetL16-17} trained on the \textsc{Kidney} data set. Right: \textsc{DVRNet} trained on the \textsc{Shaded} data set. }
    \label{fig:hd_images}
\end{figure}

In the previous experiment, we have established that the architectures are, in principle, capable of learning meaningful rendering features purely from data provided in image space. 
In this section, we perform additional experiments showing the benefits of the deep architectures we have designed.
To this end, we use a clinical data set with high inter-patient variations, which is a common scenario yet hard to define robust TFs for.


\subsubsection{Experiment setup}
The training objective is to optimize rendering models that work consistently between subjects.
The volume data consists of abdominal angiographic CT volumes (CTA) from 27 subjects cropped and resampled to an isotropic resolution $192\times 192\times 192$, together with an equally sized label volume indicating the main abdominal artery and the kidneys with separate labels. The 3D labels were exclusively used to generate the 2D training data and not at as an input to the models, as labels cannot be assumed to exist in a normal clinical setting for a new patient.
For each of the volumes, an intensity-based TF preset was adapted manually by shifting the intensity range to produce consistent images showing the bones and contrasted vessels.

From this volumetric data, we create two separate data sets via a shader-based DVR implementation: \textsc{Kidney} uses the labels to emphasize kidneys and the labeled artery by applying distinct colors and boosting the opacity. In \textsc{Shading}, we do not use the label volume but instead use an ambient and diffuse shading model with secondary shadow rays as discussed by \cite{Ropinski2008} to illuminate the ground truth images. In these images, the light source is always at the top-left of the camera (view dependent illumination), a common illumination setup for medical DVR.
For both \textsc{Kidney} and \textsc{Shading} datasets, training images are generated at a resolution of $200 \times 200$ for a fixed set of 16 viewing directions on a sphere around each volume. 
The data set was split randomly by patient, using 15 patients for training and 6 patients each for validation and testing.
Our two data sets therefore consist of 240 images in the training set and 96 images in the validation and test set, respectively.

We used the same SSIM+MSE loss function as in the previous experiment.
We have trained all networks on both data sets using the Adam\cite{Kingma2015} optimizer ($\beta_1=0.9, \beta_2 = 0.99$) and have used the same learning rate and batch size combinations as in \autoref{sec:ex_handpainted}. All models except RenderNet were trained with \emph{stepsize annealing} with $s_l=0.1, s_h=1.0$. 

\begin{figure*}
    \centering

\ra{1.0} \setlength{\tabcolsep}{0.1em}
\begin{tabular}{ >{\centering\bfseries}m{0.04\textwidth} *{6}{>{\centering}m{0.13\textwidth}}>{\centering\arraybackslash}m{0.13\textwidth}}
  & \textbf{Groundtruth} & Lookup & RenderNet & VNet4-4 & VNet16-4 & VNet16-17 & DVRNet \\ 
\cmidrule{2 - 8} 
P24 & 
    \includegraphics[width=0.13\textwidth]{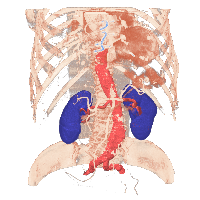} & 
    \includegraphics[width=0.13\textwidth]{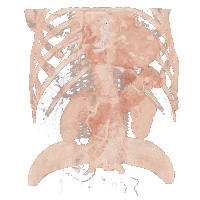} & 
    \includegraphics[width=0.13\textwidth]{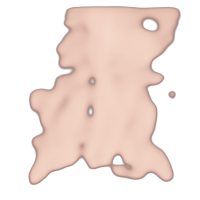} & 
    \includegraphics[width=0.13\textwidth]{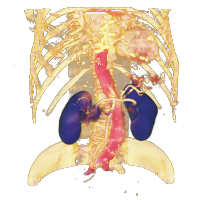} & 
    \includegraphics[width=0.13\textwidth]{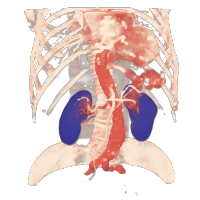} & 
    \includegraphics[width=0.13\textwidth]{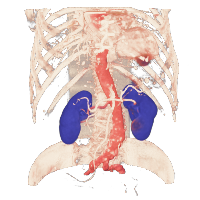} & 
    \includegraphics[width=0.13\textwidth]{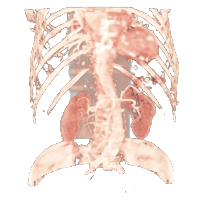} \\ 
P17 & 
    \includegraphics[width=0.13\textwidth]{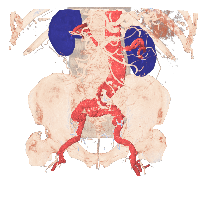} & 
    \includegraphics[width=0.13\textwidth]{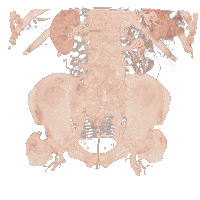} & 
    \includegraphics[width=0.13\textwidth]{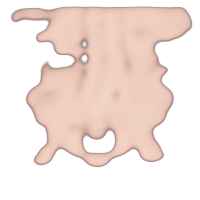} & 
    \includegraphics[width=0.13\textwidth]{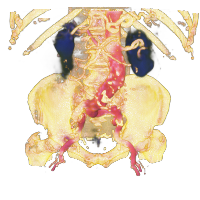} & 
    \includegraphics[width=0.13\textwidth]{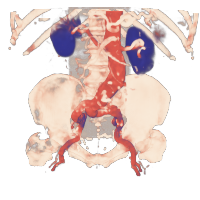} & 
    \includegraphics[width=0.13\textwidth]{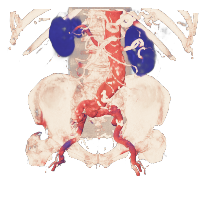} & 
    \includegraphics[width=0.13\textwidth]{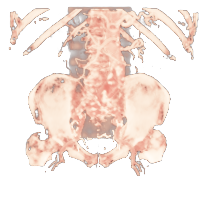} \\ 
\end{tabular}

    \caption{
    Results of our deep models trained on the \textsc{Kidney} data set. Results show the frontal view for two of the six patients from the test set. Images are best appreciated in color and at high resolution.
    }
    \label{fig:results-generalization-visual}
\end{figure*}

\begin{figure*}
    \centering
    
\ra{1.0} \setlength{\tabcolsep}{0.1em}
\begin{tabular}{ >{\centering\bfseries}m{0.04\textwidth} *{6}{>{\centering}m{0.13\textwidth}}>{\centering\arraybackslash}m{0.13\textwidth}}
  & \textbf{Groundtruth} & Lookup & RenderNet & VNet4-4 & VNet16-4 & VNet16-17 & DVRNet \\ 
\cmidrule{2 - 8} 
P24 & 
    \includegraphics[width=0.13\textwidth]{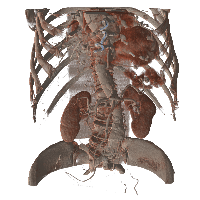} & 
    \includegraphics[width=0.13\textwidth]{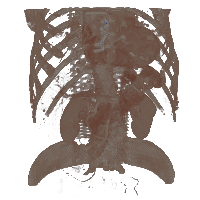} & 
    \includegraphics[width=0.13\textwidth]{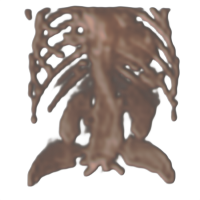} & 
    \includegraphics[width=0.13\textwidth]{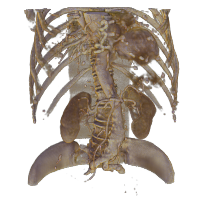} 
&
    \includegraphics[width=0.13\textwidth]{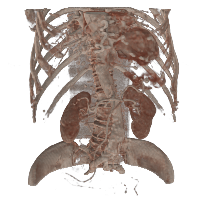} & 
    \includegraphics[width=0.13\textwidth]{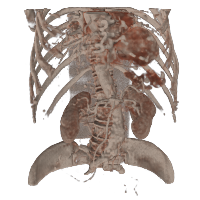} & 
    \includegraphics[width=0.13\textwidth]{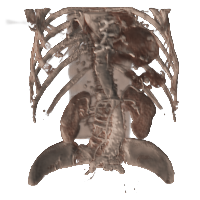} \\ 
P17 & 
    \includegraphics[width=0.13\textwidth]{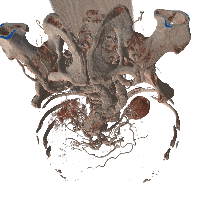} & 
    \includegraphics[width=0.13\textwidth]{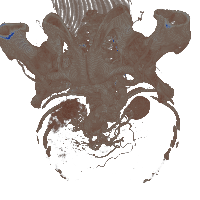} & 
    \includegraphics[width=0.13\textwidth]{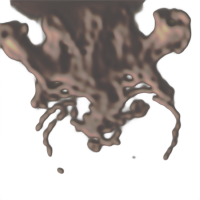} & 
    \includegraphics[width=0.13\textwidth]{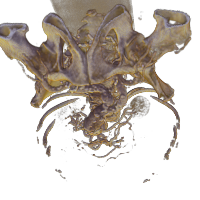} 
& 
    \includegraphics[width=0.13\textwidth]{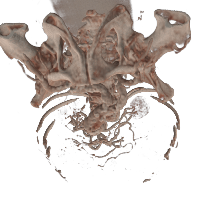} & 
    \includegraphics[width=0.13\textwidth]{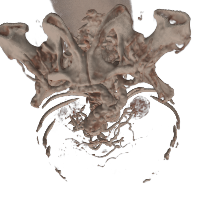} & 
    \includegraphics[width=0.13\textwidth]{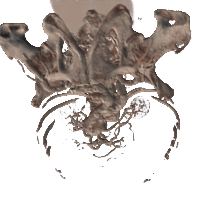} \\ 
\end{tabular}

    \caption{
    Training results for the \textsc{Kidney} data set. Results show two different viewing directions for two patients from the test set to demonstrate how the different models are able to learn view-dependent illumination.
    Images are best appreciated in color and at high resolution.
    }
    \label{fig:results-generalization-kidney}
\end{figure*}

\subsubsection{Results}
The results are summarized in \autoref{tab:generalization_results}, reporting LPIPS, FID and SSIM. However, SSIM was used in the optimization and should therefore be interpreted accordingly.
A qualitative comparison for two patients in the test set can be seen in \autoref{fig:results-generalization-visual}. Results on all six patients in the test set can be found in the supplementary document.

Inspection of the resulting images reveals that the optimized Lookup TF that was included as a baseline here misses the primary objective in both data sets: neither can it colorize the structures correctly nor is it able to represent any shading information. Both of these effects are expected since it is impossible to discriminate bone, vessel and kidney structures purely based on intensity and shading needs explicit information in view space (i.e. the angle between viewing direction and the light direction, which is relative to the camera in our data set).

Comparing the performance between the \textsc{VNet*} variants on the \textsc{Kidney} data, we can conclude that the additional channels and the trained TF of the \textsc{VNetL16-4} and \textsc{VNetL16-17} variants incrementally improve the models. Rendering the images in a latent color space of \textsc{VNetL16-17} seems to provide a slight benefit for more complex tasks as in the \textsc{Kidney} data set, however \textsc{VNet16-4} seems to produce better images on the \textsc{Shaded} training data. One reason might be that the additional parameters of \textsc{VNet16-17} are harder to optimize in this case.
On the \textsc{Kidney} data, the multiscale \textsc{DVRNet} architecture does not perform as well as the \textsc{VNet*} variants, missing both the visibility and distinct colorization in some cases.
It seems that shifting the 3D upconvolutions of VNet to 2D upconvolutions in the design of our \textsc{DVRNet} is not as effective as the full VNet when complex 3D features are required.

The 2D convolutional upsampling blocks in the decoder part however introduce the ability to perform reasoning in image space.
This advantage of \textsc{DVRNet} is visible for the \textsc{Shaded} data set, where the view-dependent illumination explicitly involves the image relative to the camera. The \textsc{VNet*} variants are not equipped for such a task as the VNet volume encoder can only learn features that are consistent between all views, and the subsequent MLPs in the more complex models can essentially only perform per-pixel calculations. Therefore, they are restricted to learn the average luminance of a spatial location seen (and illuminated) from multiple viewpoints, ending up with a result similar to ambient occlusion.
Conversely, our \textsc{DVRNet} is able to correctly and consistently learn this view-shading behavior through its multiscale image space convolutions.
Comparing qualitative and quantitative results, it seems that the perceptual scores we use in the analysis do not capture well the subtle illumination differences in this data set. SSIM, while used as part of the loss and therefore not an independent metric, shows these differences most clearly.

Notably, while RenderNet generally still produces blurry results, it seems to capture more illumination details from the \textsc{Shaded} images. This can be understood from its architecture, which transforms the volume to view space \emph{before} 3D convolutions, which helps in understanding view-dependent effects. We consider this an interesting direction for further refinement of our architectures.

Considering training time, the Lookup TF is clearly the fastest, with a total training time of around 20min. Training time of the VNet* variants are largely dominated by the deep encoder part, the added MLP TF (\textsc{VNetL16-4}) and decoder (\textsc{VNetL16-17}) only slightly increase overall training time. \textsc{DVRNet} only requires a quarter of the training time of the \textsc{VNet*} variants.

\begin{table}
\ra{1.0}
\caption{
    Results for the TF generalization task for the two data sets \textsc{Kidney} and \textsc{Shaded}.
    \textsuperscript{1}SSIM was used as part of the loss function.
}
\label{tab:results-generalization}
    \begin{minipage}{\columnwidth}
    
    \begin{center}
    \begin{tabular}{@{}rc rrrrc @{}}
    \toprule
         && \multicolumn{4}{c}{\textsc{Kidney}} \\
        \cmidrule{3-6}
    &&    LPIPS $\downarrow$ & FID $\downarrow$ & SSIM $\uparrow$\textsuperscript{1} & Train Time \\
    \midrule
    Lookup &&        0.2627 & \textbf{187.68} & \textbf{0.7060} & 22m \\
    RenderNet &&     0.5619 & 387.16 & 0.5368 & 3h 52m \\
    VNet4-4 &&       0.2683 & 280.26 & 0.6990 & 1d 2h 46m \\
    VNetL16-4 &&     0.2148 & 255.82 & 0.6936 & 1d 5h 19m \\ 
    VNetL16-17 &&    \textbf{0.2107} & 245.45 & 0.6902 & 1d 5h 26m \\
    DVRNet &&        0.2731 & 313.66 & 0.6993 & 7h 20m \\

    \midrule
    
    && \multicolumn{4}{c}{\textsc{Shaded}} \\
    \cmidrule{3-6}
    && LPIPS $\downarrow$ & FID $\downarrow$ & SSIM $\uparrow$\textsuperscript{1} & Train Time \\
    \midrule
    Lookup &&        0.2799 & 232.85 & 0.6765 & 22m \\
    RenderNet &&     0.4799 & 262.71 & 0.6218 & 3h 53m \\
    VNet4-4 &&       0.2872 & 235.57 & 0.6640 & 1d 2h 52m \\ 
    VNetL16-4 &&     \textbf{0.2405} & \textbf{225.56} & 0.6772 & 1d 5h 27m \\
    VNetL16-17 &&    0.2740 & 231.17 & 0.6561 & 1d 5h 32m \\
    DVRNet &&        0.2485 & 227.85 & \textbf{0.7337} & 7h 23m \\
        
    \bottomrule
    \end{tabular}
    \end{center}

    \end{minipage}
    \label{tab:generalization_results}
\end{table}

                \section{Discussion}
\label{sec:discussion}
We have shown in our experiments that our novel DeepDVR successfully enables end-to-end training of neural rendering architectures for volume rendering and can produce neural rendering models that achieve the desired visual outcome without manual design of input features or TFs.


Our experiments have uncovered several characteristic differences between the presented architectures:
The Lookup table provides a parametrization that remains manually adjustable through traditional user interfaces and fast training times. When only using intensity as an input feature, it is ultimately limited in its ability to discern different structures. 
Nonetheless, training a Lookup TF is useful in applications where the integration of a full neural renderer is not possible. In those cases, the lookup table can simply be transferred to existing DVR implementations. The training approach can also be readily extended to 2D TFs, where manual specification is less straightforward.

RenderNet overall is clearly outperformed by the other deep architectures we have investigated which demonstrates the effectiveness of our dedicated DVR unit.
\textsc{VNet16-17} seems to perform similar to \textsc{VNet16-4} on the TF generalization task yet it provides better results on the manually adapted reference images. 
\textsc{DVRNet} overall performed best on the manually adapted images and produced comparable renderings for the generalization tasks while requiring significantly less time for training than the VNet variants.
The specific architecture choices for \textsc{DVRNet}, involving both 3D and 2D convolutions, allow to capture both 3D spatial semantics and view-dependent aspects of the training images. However, the reduced number of input convolutions in the volume encoder make it less powerful than the full VNet-based encoders.

We want to highlight that the resampling step in the raymarching within the model architectures effectively decouples the input volume resolution from the output image resolution. Because we have chosen fully convolutional layers in the 3D and 2D portions of the architectures, our models can flexibly process arbitrary volume resolutions and produce arbitrary image resolutions. The images in \autoref{fig:hd_images} were rendered with the the two best-performing model trained in \autoref{sec:ex_generalization} but with an increased output resolution of 600$\times$600 as opposed to 200$\times$200 during training.


\paragraph{Limitations and Future Work}
It is important to acknowledge that our proposed deep architectures prevent the direct adaptation in cases where the rendering is not acceptable for a new volume. 
While removing manual interaction was the dedicated goal of our method, this is relevant until models are good enough that they "just work", so this argument merits consideration.
We could for example imagine a hybrid combination of learned 2D lookup table and learned features, where the lookup table remains editable with conventional interfaces. Perceptual loss functions\cite{Zhang2017,Johnson2016} or GANs could also be used for a final patient-specific optimization step of a pretrained model to account for per-patient variations outside of the training examples.


The high training time of the deep variants currently limits the range of applications to where precomputation can be done offline. However, the modular architecture and explicit modeling of color and opacity functions in our models enable defining selected presets or even partial retraining. Interactive refinement based on user feedback is an interesting future research direction, where we expect the explicit modeling of the DVR stages to play a crucial role.
In this work we did not focus on optimizations of the differentiable rendering algorithm itself.
At test time, our current implementation of DVRNet takes approximately 370ms per frame, whereas the VNet* variants take 1580-2008ms. An explicit implementation of the forward pass directly using shaders or GPU computing APIs together with caching the output from the volume encoder $\mathcal{E}(I)$ can provide significant speedups that will likely enable interactive visualization.
Further research is also required to reduce the memory footprint during training which will enable training on larger input volumes and output image sizes.

Beyond direct applications in rendering, we believe that the presented DeepDVR approach can also be used to train 3D segmentation networks from annotations in view space. Our \textsc{VNet4-4} model is essentially a 3D semantic segmentation model optimized by a visual loss function by projecting the 2D annotations into 3D. The experiments demonstrate that differentiable volume rendering in the training loop can be used to train semantic segmentation models from scratch only from simple 2D annotations.

The process of accumulating features along a ray from front to back is essentially an application of sequence processing and the alpha blending can be seen as a form of recurrent neural network. From this perspective, using RNN units like LSTM or GRU units could in the future replace the accumulation function $\mathcal{A}$ for learning more complex blending operations. This could be used to learn advanced implicit ray accumulation functions and produce learned "smart" compositing and illustrative rendering.


                \section{Conclusion}
\label{sec:conclusion}

In this paper, we explored the possibilities and applications of deep models modeling the DVR pipeline.
We show that with this approach, deep neural networks are able to learn the hidden semantic of what structures the user \emph{wants} to see purely from their indications in image space. This includes complex relations where the system learned to discriminate between larger and smaller vessels to only highlight the artery.
This act of inferring the 3D semantic that the user is looking for, and adapting the visualization accordingly, will be an immensely powerful tool to enable future intuitive user interaction with complex or high-dimensional data. 

We have shown that DeepDVR can be used with a traditional lookup table to optimize it from images. In experiments we show that representing the lookup table using an MLP results in additional costs in training and reduced quality of the produced images.
We have also presented several architectures that are motivated by the careful modeling of the generalized volume rendering algorithm we have introduced.
We have highlighted two practically relevant applications for DeepDVR: TF design by manual annotation in image space, and learning a generalized, robust rendering model from a set of individually fine-tuned renderings.
Furthermore, we have introduced \emph{stepsize annealing}, a method to accelerate training of raymarching models by starting at a high rendering step size in early epochs, decreasing over time to achieve similar performance as when training with a constant small step size.

We consider the DeepDVR approach as an "architectural blueprint" which can be combined with any 2D and 3D image-to-image architectures. 
Building a DeepDVR architecture that supports both good classification in volume space and good image-space reasoning could be as simple as combining a full VNet-based volume encoder with a full U-Net in the decoder. However, the biggest challenge in the future will be to find lightweight architectures that enable interactive visualization and online retraining.

The tight integration of volume rendering and deep learning methods we presented is a way to bring the fast-advancing field of deep learning into the field of scientific volume visualization.
We believe that the basic principle we demonstrated in this work has the potential as a foundation for data driven scientific volume visualization. We hope this paper will inspire future work at the intersection of deep learning and scientific volume rendering, leading to smarter and more expressive visualizations in the fields of medical and scientific volume visualization.




%
%
%
%
%

\bibliographystyle{ACM-Reference-Format}
\bibliography{references}


\begin{thebibliography}{46}


\ifx \showCODEN    \undefined \def \showCODEN     #1{\unskip}     \fi
\ifx \showDOI      \undefined \def \showDOI       #1{#1}\fi
\ifx \showISBNx    \undefined \def \showISBNx     #1{\unskip}     \fi
\ifx \showISBNxiii \undefined \def \showISBNxiii  #1{\unskip}     \fi
\ifx \showISSN     \undefined \def \showISSN      #1{\unskip}     \fi
\ifx \showLCCN     \undefined \def \showLCCN      #1{\unskip}     \fi
\ifx \shownote     \undefined \def \shownote      #1{#1}          \fi
\ifx \showarticletitle \undefined \def \showarticletitle #1{#1}   \fi
\ifx \showURL      \undefined \def \showURL       {\relax}        \fi
\providecommand\bibfield[2]{#2}
\providecommand\bibinfo[2]{#2}
\providecommand\natexlab[1]{#1}
\providecommand\showeprint[2][]{arXiv:#2}

\bibitem[\protect\citeauthoryear{Berger, Li, and Levine}{Berger
  et~al\mbox{.}}{2019}]%
        {Berger2019}
\bibfield{author}{\bibinfo{person}{Matthew Berger}, \bibinfo{person}{Jixian
  Li}, {and} \bibinfo{person}{Joshua~A Levine}.}
  \bibinfo{year}{2019}\natexlab{}.
\newblock \showarticletitle{{A Generative Model for Volume Rendering}}.
\newblock \bibinfo{journal}{\emph{IEEE Transactions on Visualization and
  Computer Graphics}} \bibinfo{volume}{25}, \bibinfo{number}{4}
  (\bibinfo{date}{apr} \bibinfo{year}{2019}), \bibinfo{pages}{1636--1650}.
\newblock
\showISSN{1077-2626}
\urldef\tempurl%
\url{https://doi.org/10.1109/TVCG.2018.2816059}
\showDOI{\tempurl}
\showeprint[arxiv]{1710.09545}


\bibitem[\protect\citeauthoryear{Bruckner and Gr{\"{o}}ller}{Bruckner and
  Gr{\"{o}}ller}{2009}]%
        {Bruckner2009}
\bibfield{author}{\bibinfo{person}{Stefan Bruckner} {and}
  \bibinfo{person}{M~Eduard Gr{\"{o}}ller}.} \bibinfo{year}{2009}\natexlab{}.
\newblock \showarticletitle{{Instant Volume Visualization using Maximum
  Intensity Difference Accumulation}}.
\newblock \bibinfo{journal}{\emph{Computer Graphics Forum}}
  \bibinfo{volume}{28}, \bibinfo{number}{3} (\bibinfo{year}{2009}),
  \bibinfo{pages}{775--782}.
\newblock
\showISBNx{01677055 14678659}
\urldef\tempurl%
\url{https://doi.org/10.1111/j.1467-8659.2009.01474.x}
\showDOI{\tempurl}


\bibitem[\protect\citeauthoryear{Correa and {Kwan-Liu Ma}}{Correa and {Kwan-Liu
  Ma}}{2008}]%
        {Correa2008a}
\bibfield{author}{\bibinfo{person}{C. Correa} {and} \bibinfo{person}{{Kwan-Liu
  Ma}}.} \bibinfo{year}{2008}\natexlab{}.
\newblock \showarticletitle{{Size-based Transfer Functions: A New Volume
  Exploration Technique}}.
\newblock \bibinfo{journal}{\emph{IEEE Transactions on Visualization and
  Computer Graphics}} \bibinfo{volume}{14}, \bibinfo{number}{6}
  (\bibinfo{date}{nov} \bibinfo{year}{2008}), \bibinfo{pages}{1380--1387}.
\newblock
\showISSN{1077-2626}
\urldef\tempurl%
\url{https://doi.org/10.1109/TVCG.2008.162}
\showDOI{\tempurl}


\bibitem[\protect\citeauthoryear{Correa and Ma}{Correa and Ma}{2009}]%
        {Correa2009}
\bibfield{author}{\bibinfo{person}{Carlos~D. Correa} {and}
  \bibinfo{person}{Kwan~Liu Ma}.} \bibinfo{year}{2009}\natexlab{}.
\newblock \showarticletitle{{The occlusion spectrum for volume classification
  and visualization}}.
\newblock \bibinfo{journal}{\emph{IEEE Transactions on Visualization and
  Computer Graphics}} \bibinfo{volume}{15}, \bibinfo{number}{6}
  (\bibinfo{year}{2009}), \bibinfo{pages}{1465--1472}.
\newblock
\showISBNx{1077-2626}
\showISSN{10772626}
\urldef\tempurl%
\url{https://doi.org/10.1109/TVCG.2009.189}
\showDOI{\tempurl}


\bibitem[\protect\citeauthoryear{Danskin and Hanrahan}{Danskin and
  Hanrahan}{1992}]%
        {Danskin1992}
\bibfield{author}{\bibinfo{person}{John Danskin} {and} \bibinfo{person}{Pat
  Hanrahan}.} \bibinfo{year}{1992}\natexlab{}.
\newblock \showarticletitle{{Fast algorithms for volume ray tracing}}. In
  \bibinfo{booktitle}{\emph{Proceedings of the 1992 workshop on Volume
  visualization - VVS '92}}. \bibinfo{publisher}{ACM Press},
  \bibinfo{address}{New York, New York, USA}, \bibinfo{pages}{91--98}.
\newblock
\showISBNx{0897915275}
\urldef\tempurl%
\url{https://doi.org/10.1145/147130.147155}
\showDOI{\tempurl}


\bibitem[\protect\citeauthoryear{{De Moura Pinto} and Freitas}{{De Moura Pinto}
  and Freitas}{2007}]%
        {DeMouraPinto2007}
\bibfield{author}{\bibinfo{person}{Francisco {De Moura Pinto}} {and}
  \bibinfo{person}{Carla~M.D.S. Freitas}.} \bibinfo{year}{2007}\natexlab{}.
\newblock \showarticletitle{{Design of Multi-dimensional Transfer Functions
  Using Dimensional Reduction}}.
\newblock \bibinfo{journal}{\emph{Eurographics/IEEE-VGTC Symposium on
  Visualization}} \bibinfo{number}{February} (\bibinfo{year}{2007}).
\newblock
\urldef\tempurl%
\url{https://doi.org/10.2312/VisSym/EuroVis07/131-138}
\showDOI{\tempurl}


\bibitem[\protect\citeauthoryear{{de Moura Pinto} and Freitas}{{de Moura Pinto}
  and Freitas}{2010}]%
        {DeMouraPinto2010a}
\bibfield{author}{\bibinfo{person}{Francisco {de Moura Pinto}} {and}
  \bibinfo{person}{C~M D~S Freitas}.} \bibinfo{year}{2010}\natexlab{}.
\newblock \showarticletitle{{Importance-Aware Composition for Illustrative
  Volume Rendering}}. In \bibinfo{booktitle}{\emph{2010 23rd SIBGRAPI
  Conference on Graphics, Patterns and Images}}. \bibinfo{publisher}{IEEE},
  \bibinfo{pages}{134--141}.
\newblock
\showISBNx{978-1-4244-8420-1}
\urldef\tempurl%
\url{https://doi.org/10.1109/SIBGRAPI.2010.26}
\showDOI{\tempurl}


\bibitem[\protect\citeauthoryear{Engel and Ropinski}{Engel and
  Ropinski}{2020}]%
        {Engel2020}
\bibfield{author}{\bibinfo{person}{Dominik Engel} {and} \bibinfo{person}{Timo
  Ropinski}.} \bibinfo{year}{2020}\natexlab{}.
\newblock \showarticletitle{{Deep Volumetric Ambient Occlusion}}.
\newblock  (\bibinfo{date}{aug} \bibinfo{year}{2020}).
\newblock
\showeprint[arxiv]{2008.08345}
\urldef\tempurl%
\url{https://github.com/xeTaiz/dvao http://arxiv.org/abs/2008.08345}
\showURL{%
\tempurl}


\bibitem[\protect\citeauthoryear{Engel, Kraus, and Ertl}{Engel
  et~al\mbox{.}}{2001}]%
        {Engel2001}
\bibfield{author}{\bibinfo{person}{Klaus Engel}, \bibinfo{person}{Martin
  Kraus}, {and} \bibinfo{person}{Thomas Ertl}.}
  \bibinfo{year}{2001}\natexlab{}.
\newblock \showarticletitle{{High-quality pre-integrated volume rendering using
  hardware-accelerated pixel shading}}. In
  \bibinfo{booktitle}{\emph{Proceedings of the ACM SIGGRAPH/EUROGRAPHICS
  workshop on Graphics hardware - HWWS '01}}. \bibinfo{publisher}{ACM Press},
  \bibinfo{address}{New York, New York, USA}, \bibinfo{pages}{9--16}.
\newblock
\showISBNx{158113407X}
\urldef\tempurl%
\url{https://doi.org/10.1145/383507.383515}
\showDOI{\tempurl}


\bibitem[\protect\citeauthoryear{Hadwiger, Ljung, Salama, and
  Ropinski}{Hadwiger et~al\mbox{.}}{2008}]%
        {Hadwiger2008a}
\bibfield{author}{\bibinfo{person}{Markus Hadwiger}, \bibinfo{person}{Patric
  Ljung}, \bibinfo{person}{Christof~Rezk Salama}, {and} \bibinfo{person}{Timo
  Ropinski}.} \bibinfo{year}{2008}\natexlab{}.
\newblock \showarticletitle{{Advanced illumination techniques for GPU volume
  raycasting}}.
\newblock \bibinfo{journal}{\emph{ACM SIGGRAPH ASIA 2008 Courses}}
  (\bibinfo{year}{2008}), \bibinfo{pages}{1--166}.
\newblock
\showISBNx{2717403337}
\urldef\tempurl%
\url{https://doi.org/10.1145/1508044.1508045}
\showDOI{\tempurl}


\bibitem[\protect\citeauthoryear{{Hanqi Guo}, {Ningyu Mao}, and {Xiaoru
  Yuan}}{{Hanqi Guo} et~al\mbox{.}}{2011}]%
        {HanqiGuo2011}
\bibfield{author}{\bibinfo{person}{{Hanqi Guo}}, \bibinfo{person}{{Ningyu
  Mao}}, {and} \bibinfo{person}{{Xiaoru Yuan}}.}
  \bibinfo{year}{2011}\natexlab{}.
\newblock \showarticletitle{{WYSIWYG (What You See is What You Get) Volume
  Visualization}}.
\newblock \bibinfo{journal}{\emph{IEEE Transactions on Visualization and
  Computer Graphics}} \bibinfo{volume}{17}, \bibinfo{number}{12}
  (\bibinfo{date}{dec} \bibinfo{year}{2011}), \bibinfo{pages}{2106--2114}.
\newblock
\showISBNx{978-1-7281-5697-2}
\showISSN{1077-2626}
\urldef\tempurl%
\url{https://doi.org/10.1109/TVCG.2011.261}
\showDOI{\tempurl}


\bibitem[\protect\citeauthoryear{He, Wang, Guo, Wang, Shen, Raj, Nashed, and
  Peterka}{He et~al\mbox{.}}{2020}]%
        {He2020}
\bibfield{author}{\bibinfo{person}{Wenbin He}, \bibinfo{person}{Junpeng Wang},
  \bibinfo{person}{Hanqi Guo}, \bibinfo{person}{Ko~Chih Wang},
  \bibinfo{person}{Han~Wei Shen}, \bibinfo{person}{Mukund Raj},
  \bibinfo{person}{Youssef~S.G. Nashed}, {and} \bibinfo{person}{Tom Peterka}.}
  \bibinfo{year}{2020}\natexlab{}.
\newblock \showarticletitle{{InSituNet: Deep Image Synthesis for Parameter
  Space Exploration of Ensemble Simulations}}.
\newblock \bibinfo{journal}{\emph{IEEE Transactions on Visualization and
  Computer Graphics}} \bibinfo{volume}{26}, \bibinfo{number}{1}
  (\bibinfo{year}{2020}), \bibinfo{pages}{23--33}.
\newblock
\showISSN{19410506}
\urldef\tempurl%
\url{https://doi.org/10.1109/TVCG.2019.2934312}
\showDOI{\tempurl}
\showeprint[arxiv]{1908.00407}


\bibitem[\protect\citeauthoryear{Heusel, Ramsauer, Unterthiner, Nessler, and
  Hochreiter}{Heusel et~al\mbox{.}}{2017}]%
        {Heusel2017}
\bibfield{author}{\bibinfo{person}{Martin Heusel}, \bibinfo{person}{Hubert
  Ramsauer}, \bibinfo{person}{Thomas Unterthiner}, \bibinfo{person}{Bernhard
  Nessler}, {and} \bibinfo{person}{Sepp Hochreiter}.}
  \bibinfo{year}{2017}\natexlab{}.
\newblock \showarticletitle{{GANs trained by a two time-scale update rule
  converge to a local Nash equilibrium}}. In \bibinfo{booktitle}{\emph{Advances
  in Neural Information Processing Systems}},
  Vol.~\bibinfo{volume}{2017-Decem}. \bibinfo{pages}{6627--6638}.
\newblock
\showISSN{10495258}


\bibitem[\protect\citeauthoryear{Hong, Liu, and Yuan}{Hong
  et~al\mbox{.}}{2019}]%
        {Hong2019}
\bibfield{author}{\bibinfo{person}{Fan Hong}, \bibinfo{person}{Can Liu}, {and}
  \bibinfo{person}{Xiaoru Yuan}.} \bibinfo{year}{2019}\natexlab{}.
\newblock \showarticletitle{{DNN-VolVis: Interactive volume visualization
  supported by deep neural network}}.
\newblock \bibinfo{journal}{\emph{IEEE Pacific Visualization Symposium}}
  \bibinfo{volume}{2019-April} (\bibinfo{year}{2019}),
  \bibinfo{pages}{282--291}.
\newblock
\showISBNx{9781538692264}
\showISSN{21658773}
\urldef\tempurl%
\url{https://doi.org/10.1109/PacificVis.2019.00041}
\showDOI{\tempurl}


\bibitem[\protect\citeauthoryear{Jain, Griffin, Godil, Bullard, Terrill, and
  Varshney}{Jain et~al\mbox{.}}{2017}]%
        {Jain2017}
\bibfield{author}{\bibinfo{person}{Somay Jain}, \bibinfo{person}{Wesley
  Griffin}, \bibinfo{person}{Afzal Godil}, \bibinfo{person}{Jeffrey~W Bullard},
  \bibinfo{person}{Judith Terrill}, {and} \bibinfo{person}{Amitabh Varshney}.}
  \bibinfo{year}{2017}\natexlab{}.
\newblock \showarticletitle{{Compressed Volume Rendering using Deep Learning}}.
  In \bibinfo{booktitle}{\emph{Proceedings of the Large Scale Data Analysis and
  Visualization (LDAV) Symposium}}. \bibinfo{address}{Phoenix, AZ}.
\newblock


\bibitem[\protect\citeauthoryear{Johnson, Alahi, and Fei-Fei}{Johnson
  et~al\mbox{.}}{2016}]%
        {Johnson2016}
\bibfield{author}{\bibinfo{person}{Justin Johnson}, \bibinfo{person}{Alexandre
  Alahi}, {and} \bibinfo{person}{Li Fei-Fei}.} \bibinfo{year}{2016}\natexlab{}.
\newblock \showarticletitle{{Perceptual losses for real-time style transfer and
  super-resolution}}. In \bibinfo{booktitle}{\emph{Lecture Notes in Computer
  Science (including subseries Lecture Notes in Artificial Intelligence and
  Lecture Notes in Bioinformatics)}}, Vol.~\bibinfo{volume}{9906 LNCS}.
  \bibinfo{pages}{694--711}.
\newblock
\showISBNx{9783319464749}
\showISSN{16113349}
\urldef\tempurl%
\url{https://doi.org/10.1007/978-3-319-46475-6_43}
\showDOI{\tempurl}
\showeprint[arxiv]{1603.08155v1}


\bibitem[\protect\citeauthoryear{Kato, Beker, Morariu, Ando, Matsuoka, Kehl,
  and Gaidon}{Kato et~al\mbox{.}}{2020}]%
        {Kato2020}
\bibfield{author}{\bibinfo{person}{Hiroharu Kato}, \bibinfo{person}{Deniz
  Beker}, \bibinfo{person}{Mihai Morariu}, \bibinfo{person}{Takahiro Ando},
  \bibinfo{person}{Toru Matsuoka}, \bibinfo{person}{Wadim Kehl}, {and}
  \bibinfo{person}{Adrien Gaidon}.} \bibinfo{year}{2020}\natexlab{}.
\newblock \showarticletitle{{Differentiable Rendering: A Survey}}.
\newblock  (\bibinfo{date}{jun} \bibinfo{year}{2020}).
\newblock
\showeprint[arxiv]{2006.12057}
\urldef\tempurl%
\url{http://arxiv.org/abs/2006.12057}
\showURL{%
\tempurl}


\bibitem[\protect\citeauthoryear{Kindlmann, Whitaker, Tasdizen, and
  Moller}{Kindlmann et~al\mbox{.}}{2003}]%
        {Kindlmann2003}
\bibfield{author}{\bibinfo{person}{G Kindlmann}, \bibinfo{person}{R Whitaker},
  \bibinfo{person}{T Tasdizen}, {and} \bibinfo{person}{T Moller}.}
  \bibinfo{year}{2003}\natexlab{}.
\newblock \showarticletitle{{Curvature-based transfer functions for direct
  volume rendering: methods and applications}}.
\newblock  (\bibinfo{year}{2003}), \bibinfo{pages}{513--520}.
\newblock
\urldef\tempurl%
\url{https://doi.org/10.1109/visual.2003.1250414}
\showDOI{\tempurl}


\bibitem[\protect\citeauthoryear{Kingma and Ba}{Kingma and Ba}{2015}]%
        {Kingma2015}
\bibfield{author}{\bibinfo{person}{Diederik~P Kingma} {and}
  \bibinfo{person}{Jimmy~Lei Ba}.} \bibinfo{year}{2015}\natexlab{}.
\newblock \showarticletitle{{Adam: A method for stochastic optimization}}. In
  \bibinfo{booktitle}{\emph{3rd International Conference on Learning
  Representations, ICLR 2015 - Conference Track Proceedings}}.
\newblock
\showeprint[arxiv]{1412.6980}


\bibitem[\protect\citeauthoryear{Kniss, Kindlmann, and Hansen}{Kniss
  et~al\mbox{.}}{2001}]%
        {Kniss2001}
\bibfield{author}{\bibinfo{person}{Joe Kniss}, \bibinfo{person}{Gordon
  Kindlmann}, {and} \bibinfo{person}{Charles Hansen}.}
  \bibinfo{year}{2001}\natexlab{}.
\newblock \showarticletitle{{Interactive volume rendering using
  multi-dimensional transfer functions and direct manipulation widgets}}. In
  \bibinfo{booktitle}{\emph{Proceedings Visualization, 2001. VIS '01.}}
  \bibinfo{publisher}{IEEE}, \bibinfo{pages}{255--562}.
\newblock
\showISBNx{0-7803-7200-X}
\urldef\tempurl%
\url{https://doi.org/10.1109/VISUAL.2001.964519}
\showDOI{\tempurl}


\bibitem[\protect\citeauthoryear{Kniss, Premo{\v{z}}e, Ikits, Lefohn, Hansen,
  and Praun}{Kniss et~al\mbox{.}}{2003}]%
        {Kniss2003}
\bibfield{author}{\bibinfo{person}{Joe Kniss}, \bibinfo{person}{Simon
  Premo{\v{z}}e}, \bibinfo{person}{Milan Ikits}, \bibinfo{person}{Aaron
  Lefohn}, \bibinfo{person}{Charles Hansen}, {and} \bibinfo{person}{Emil
  Praun}.} \bibinfo{year}{2003}\natexlab{}.
\newblock \showarticletitle{{Gaussian Transfer Functions for Multi-Field Volume
  Visualization}}.
\newblock \bibinfo{journal}{\emph{Proceedings of the IEEE Visualization
  Conference}}  \bibinfo{volume}{Vi} (\bibinfo{year}{2003}),
  \bibinfo{pages}{497--504}.
\newblock
\showISBNx{0780381203}
\urldef\tempurl%
\url{https://doi.org/10.1109/VISUAL.2003.1250412}
\showDOI{\tempurl}


\bibitem[\protect\citeauthoryear{Kr{\"{u}}ger and Westermann}{Kr{\"{u}}ger and
  Westermann}{2004}]%
        {Kruger2003}
\bibfield{author}{\bibinfo{person}{J. Kr{\"{u}}ger} {and} \bibinfo{person}{R.
  Westermann}.} \bibinfo{year}{2004}\natexlab{}.
\newblock \showarticletitle{{Acceleration techniques for GPU-based volume
  rendering}}.
\newblock \bibinfo{journal}{\emph{Proceedings of the 14th IEEE Visualization
  2003 (VIS'03)}} (\bibinfo{year}{2004}), \bibinfo{pages}{287--292}.
\newblock
\urldef\tempurl%
\url{https://doi.org/10.1109/visual.2003.1250384}
\showDOI{\tempurl}


\bibitem[\protect\citeauthoryear{Li, Aittala, Durand, and Lehtinen}{Li
  et~al\mbox{.}}{2018}]%
        {Li2018}
\bibfield{author}{\bibinfo{person}{Tzu~Mao Li}, \bibinfo{person}{Miika
  Aittala}, \bibinfo{person}{Fr{\'{e}}do Durand}, {and} \bibinfo{person}{Jaakko
  Lehtinen}.} \bibinfo{year}{2018}\natexlab{}.
\newblock \showarticletitle{{Differentiable Monte Carlo ray tracing through
  edge sampling}}. In \bibinfo{booktitle}{\emph{SIGGRAPH Asia 2018 Technical
  Papers, SIGGRAPH Asia 2018}}.
\newblock
\showISBNx{9781450360081}
\showISSN{15577368}
\urldef\tempurl%
\url{https://doi.org/10.1145/3272127.3275109}
\showDOI{\tempurl}


\bibitem[\protect\citeauthoryear{Liu, Chen, Li, and Li}{Liu
  et~al\mbox{.}}{2019}]%
        {Liu2019}
\bibfield{author}{\bibinfo{person}{Shichen Liu}, \bibinfo{person}{Weikai Chen},
  \bibinfo{person}{Tianye Li}, {and} \bibinfo{person}{Hao Li}.}
  \bibinfo{year}{2019}\natexlab{}.
\newblock \showarticletitle{{Soft rasterizer: A differentiable renderer for
  image-based 3D reasoning}}. In \bibinfo{booktitle}{\emph{Proceedings of the
  IEEE International Conference on Computer Vision}},
  Vol.~\bibinfo{volume}{2019-Octob}. \bibinfo{pages}{7707--7716}.
\newblock
\showISBNx{9781728148038}
\showISSN{15505499}
\urldef\tempurl%
\url{https://doi.org/10.1109/ICCV.2019.00780}
\showDOI{\tempurl}
\showeprint[arxiv]{1904.01786}


\bibitem[\protect\citeauthoryear{Liu, Zhang, Peng, Shi, Pollefeys, and Cui}{Liu
  et~al\mbox{.}}{2020}]%
        {Liu2020}
\bibfield{author}{\bibinfo{person}{Shaohui Liu}, \bibinfo{person}{Yinda Zhang},
  \bibinfo{person}{Songyou Peng}, \bibinfo{person}{Boxin Shi},
  \bibinfo{person}{Marc Pollefeys}, {and} \bibinfo{person}{Zhaopeng Cui}.}
  \bibinfo{year}{2020}\natexlab{}.
\newblock \showarticletitle{{DIST: Rendering Deep Implicit Signed Distance
  Function With Differentiable Sphere Tracing}}. In
  \bibinfo{booktitle}{\emph{2020 IEEE/CVF Conference on Computer Vision and
  Pattern Recognition (CVPR)}}. \bibinfo{publisher}{IEEE},
  \bibinfo{pages}{2016--2025}.
\newblock
\showISBNx{978-1-7281-7168-5}
\urldef\tempurl%
\url{https://doi.org/10.1109/CVPR42600.2020.00209}
\showDOI{\tempurl}
\showeprint[arxiv]{1911.13225}


\bibitem[\protect\citeauthoryear{Ljung, Kr{\"{u}}ger, Groller, Hadwiger,
  Hansen, and Ynnerman}{Ljung et~al\mbox{.}}{2016}]%
        {Ljung2016}
\bibfield{author}{\bibinfo{person}{Patric Ljung}, \bibinfo{person}{Jens
  Kr{\"{u}}ger}, \bibinfo{person}{Eduard Groller}, \bibinfo{person}{Markus
  Hadwiger}, \bibinfo{person}{Charles~D Hansen}, {and} \bibinfo{person}{Anders
  Ynnerman}.} \bibinfo{year}{2016}\natexlab{}.
\newblock \showarticletitle{{State of the Art in Transfer Functions for Direct
  Volume Rendering}}.
\newblock \bibinfo{journal}{\emph{Computer Graphics Forum}}
  \bibinfo{volume}{35}, \bibinfo{number}{3} (\bibinfo{year}{2016}),
  \bibinfo{pages}{669--691}.
\newblock
\showISBNx{01677055}
\urldef\tempurl%
\url{https://doi.org/10.1111/cgf.12934}
\showDOI{\tempurl}


\bibitem[\protect\citeauthoryear{Lombardi, Simon, Saragih, Schwartz, Lehrmann,
  and Sheikh}{Lombardi et~al\mbox{.}}{2019}]%
        {Lombardi2019b}
\bibfield{author}{\bibinfo{person}{Stephen Lombardi}, \bibinfo{person}{Tomas
  Simon}, \bibinfo{person}{Jason Saragih}, \bibinfo{person}{Gabriel Schwartz},
  \bibinfo{person}{Andreas Lehrmann}, {and} \bibinfo{person}{Yaser Sheikh}.}
  \bibinfo{year}{2019}\natexlab{}.
\newblock \showarticletitle{{Neural volumes: Learning dynamic renderable
  volumes from images}}.
\newblock \bibinfo{journal}{\emph{ACM Transactions on Graphics}}
  \bibinfo{volume}{38}, \bibinfo{number}{4} (\bibinfo{year}{2019}),
  \bibinfo{pages}{14}.
\newblock
\showISSN{15577368}
\urldef\tempurl%
\url{https://doi.org/10.1145/3306346.3323020}
\showDOI{\tempurl}
\showeprint[arxiv]{1906.07751}


\bibitem[\protect\citeauthoryear{Loubet, Holzschuch, and Jakob}{Loubet
  et~al\mbox{.}}{2019}]%
        {Loubet2019}
\bibfield{author}{\bibinfo{person}{Guillaume Loubet}, \bibinfo{person}{Nicolas
  Holzschuch}, {and} \bibinfo{person}{Wenzel Jakob}.}
  \bibinfo{year}{2019}\natexlab{}.
\newblock \showarticletitle{{Reparame-terizing Discontinuous Integrands for
  Differentiable Rendering}}.
\newblock \bibinfo{journal}{\emph{ACM Trans. Graph}}  \bibinfo{volume}{38}
  (\bibinfo{year}{2019}), \bibinfo{pages}{14}.
\newblock
\urldef\tempurl%
\url{https://doi.org/10.1145/3355089.3356510}
\showDOI{\tempurl}


\bibitem[\protect\citeauthoryear{Max}{Max}{1995}]%
        {Max1995}
\bibfield{author}{\bibinfo{person}{Nelson Max}.}
  \bibinfo{year}{1995}\natexlab{}.
\newblock \showarticletitle{{Optical models for direct volume rendering}}.
\newblock \bibinfo{journal}{\emph{IEEE Transactions on Visualization and
  Computer Graphics}} \bibinfo{volume}{1}, \bibinfo{number}{2}
  (\bibinfo{date}{jun} \bibinfo{year}{1995}), \bibinfo{pages}{99--108}.
\newblock
\showISBNx{1077-2626}
\showISSN{10772626}
\urldef\tempurl%
\url{https://doi.org/10.1109/2945.468400}
\showDOI{\tempurl}


\bibitem[\protect\citeauthoryear{Mildenhall, Srinivasan, Tancik, Barron,
  Ramamoorthi, and Ng}{Mildenhall et~al\mbox{.}}{2020}]%
        {Mildenhall2020a}
\bibfield{author}{\bibinfo{person}{Ben Mildenhall}, \bibinfo{person}{Pratul~P
  Srinivasan}, \bibinfo{person}{Matthew Tancik}, \bibinfo{person}{Jonathan~T
  Barron}, \bibinfo{person}{Ravi Ramamoorthi}, {and} \bibinfo{person}{Ren Ng}.}
  \bibinfo{year}{2020}\natexlab{}.
\newblock \showarticletitle{{NeRF: Representing Scenes as Neural Radiance
  Fields for View Synthesis}}.
\newblock  (\bibinfo{date}{mar} \bibinfo{year}{2020}).
\newblock
\showeprint[arxiv]{2003.08934}
\urldef\tempurl%
\url{http://arxiv.org/abs/2003.08934}
\showURL{%
\tempurl}


\bibitem[\protect\citeauthoryear{Milletari, Navab, and Ahmadi}{Milletari
  et~al\mbox{.}}{2016}]%
        {Milletari2016}
\bibfield{author}{\bibinfo{person}{Fausto Milletari}, \bibinfo{person}{Nassir
  Navab}, {and} \bibinfo{person}{Seyed-Ahmad Ahmadi}.}
  \bibinfo{year}{2016}\natexlab{}.
\newblock \showarticletitle{{V-Net: Fully Convolutional Neural Networks for
  Volumetric Medical Image Segmentation}}. In \bibinfo{booktitle}{\emph{2016
  Fourth International Conference on 3D Vision (3DV)}}.
  \bibinfo{publisher}{IEEE}, \bibinfo{pages}{565--571}.
\newblock
\showISBNx{978-1-5090-5407-7}
\urldef\tempurl%
\url{https://doi.org/10.1109/3DV.2016.79}
\showDOI{\tempurl}
\showeprint[arxiv]{1606.04797v1}


\bibitem[\protect\citeauthoryear{Nguyen-Phuoc, Li, Yang, and
  Balaban}{Nguyen-Phuoc et~al\mbox{.}}{2018}]%
        {Nguyen-Phuoc2018}
\bibfield{author}{\bibinfo{person}{Thu Nguyen-Phuoc}, \bibinfo{person}{Chuan
  Li}, \bibinfo{person}{Yong~Liang Yang}, {and} \bibinfo{person}{Stephen
  Balaban}.} \bibinfo{year}{2018}\natexlab{}.
\newblock \showarticletitle{{Rendernet: A deep convolutional network for
  differentiable rendering from 3D shapes}}. In
  \bibinfo{booktitle}{\emph{Advances in Neural Information Processing
  Systems}}. \bibinfo{pages}{7891--7901}.
\newblock
\showISSN{10495258}
\showeprint[arxiv]{1806.06575}


\bibitem[\protect\citeauthoryear{Niemeyer, Mescheder, Oechsle, and
  Geiger}{Niemeyer et~al\mbox{.}}{2020}]%
        {Niemeyer2020}
\bibfield{author}{\bibinfo{person}{Michael Niemeyer}, \bibinfo{person}{Lars
  Mescheder}, \bibinfo{person}{Michael Oechsle}, {and} \bibinfo{person}{Andreas
  Geiger}.} \bibinfo{year}{2020}\natexlab{}.
\newblock \showarticletitle{{Differentiable Volumetric Rendering: Learning
  Implicit 3D Representations Without 3D Supervision}}. In
  \bibinfo{booktitle}{\emph{2020 IEEE/CVF Conference on Computer Vision and
  Pattern Recognition (CVPR)}}. \bibinfo{publisher}{IEEE},
  \bibinfo{pages}{3501--3512}.
\newblock
\showISBNx{978-1-7281-7168-5}
\urldef\tempurl%
\url{https://doi.org/10.1109/CVPR42600.2020.00356}
\showDOI{\tempurl}
\showeprint[arxiv]{1912.07372}


\bibitem[\protect\citeauthoryear{Nimier-David, Polytechnique, {De Lausanne},
  Vicini, Zeltner, and Jakob}{Nimier-David et~al\mbox{.}}{2019}]%
        {Nimier-David2019}
\bibfield{author}{\bibinfo{person}{Merlin Nimier-David},
  \bibinfo{person}{{\'{E}}cole Polytechnique},
  \bibinfo{person}{F{\'{e}}d{\'{e}}rale {De Lausanne}}, \bibinfo{person}{Delio
  Vicini}, \bibinfo{person}{Tizian Zeltner}, {and} \bibinfo{person}{Wenzel
  Jakob}.} \bibinfo{year}{2019}\natexlab{}.
\newblock \showarticletitle{{Mitsuba 2: A Retargetable Forward and Inverse
  Renderer}}.
\newblock \bibinfo{journal}{\emph{ACM Trans. Graph}} \bibinfo{volume}{38},
  \bibinfo{number}{6} (\bibinfo{year}{2019}), \bibinfo{pages}{17}.
\newblock
\urldef\tempurl%
\url{https://doi.org/10.1145/3355089.3356498}
\showDOI{\tempurl}


\bibitem[\protect\citeauthoryear{Paszke, Gross, Chintala, Chanan, Yang, and
  ...}{Paszke et~al\mbox{.}}{2017}]%
        {Paszke2017}
\bibfield{author}{\bibinfo{person}{Adam Paszke}, \bibinfo{person}{Sam Gross},
  \bibinfo{person}{Soumith Chintala}, \bibinfo{person}{Gregory Chanan},
  \bibinfo{person}{Edward Yang}, {and} \bibinfo{person}{...}}
  \bibinfo{year}{2017}\natexlab{}.
\newblock \showarticletitle{{Automatic differentiation in pytorch}}. In
  \bibinfo{booktitle}{\emph{NIPS 2017 Workshop Autodiff}}.
\newblock
\urldef\tempurl%
\url{https://openreview.net/forum?id=BJJsrmfCZ}
\showURL{%
\tempurl}


\bibitem[\protect\citeauthoryear{Rematas and Ferrari}{Rematas and
  Ferrari}{2020}]%
        {Rematas2020}
\bibfield{author}{\bibinfo{person}{Konstantinos Rematas} {and}
  \bibinfo{person}{Vittorio Ferrari}.} \bibinfo{year}{2020}\natexlab{}.
\newblock \showarticletitle{{Neural Voxel Renderer: Learning an Accurate and
  Controllable Rendering Tool}}. In \bibinfo{booktitle}{\emph{2020 IEEE/CVF
  Conference on Computer Vision and Pattern Recognition (CVPR)}}.
  \bibinfo{publisher}{IEEE}, \bibinfo{pages}{5416--5426}.
\newblock
\showISBNx{978-1-7281-7168-5}
\urldef\tempurl%
\url{https://doi.org/10.1109/CVPR42600.2020.00546}
\showDOI{\tempurl}
\showeprint[arxiv]{1912.04591}


\bibitem[\protect\citeauthoryear{Rezk-Salama, Keller, and Kohlmann}{Rezk-Salama
  et~al\mbox{.}}{2006}]%
        {Rezk-Salama2006}
\bibfield{author}{\bibinfo{person}{Christof Rezk-Salama}, \bibinfo{person}{Maik
  Keller}, {and} \bibinfo{person}{Peter Kohlmann}.}
  \bibinfo{year}{2006}\natexlab{}.
\newblock \showarticletitle{{High-Level User Interfaces for Transfer Function
  Design with Semantics}}.
\newblock \bibinfo{journal}{\emph{IEEE Transactions on Visualization and
  Computer Graphics}} \bibinfo{volume}{12}, \bibinfo{number}{5}
  (\bibinfo{date}{sep} \bibinfo{year}{2006}), \bibinfo{pages}{1021--1028}.
\newblock
\showISSN{1077-2626}
\urldef\tempurl%
\url{https://doi.org/10.1109/TVCG.2006.148}
\showDOI{\tempurl}


\bibitem[\protect\citeauthoryear{Ropinski, Pra{\ss}ni, Steinicke, and
  Hinrichs}{Ropinski et~al\mbox{.}}{2008}]%
        {Ropinski2008}
\bibfield{author}{\bibinfo{person}{Timo Ropinski}, \bibinfo{person}{J
  Pra{\ss}ni}, \bibinfo{person}{Frank Steinicke}, {and} \bibinfo{person}{Klaus
  Hinrichs}.} \bibinfo{year}{2008}\natexlab{}.
\newblock \showarticletitle{{Stroke-based transfer function design}}.
\newblock \bibinfo{journal}{\emph{Proceedings of the Fifth Eurographics/IEEE
  VGTC conference on Point-Based Graphics}} (\bibinfo{year}{2008}),
  \bibinfo{pages}{41--48}.
\newblock
\showISBNx{3905674122}
\urldef\tempurl%
\url{https://basilic.informatik.uni-hamburg.de/Publications/2008/RPSH08/vg08.pdf}
\showURL{%
\tempurl}


\bibitem[\protect\citeauthoryear{R{\"{o}}ttger}{R{\"{o}}ttger}{2020}]%
        {Rottger2020}
\bibfield{author}{\bibinfo{person}{Stefan R{\"{o}}ttger}.}
  \bibinfo{year}{2020}\natexlab{}.
\newblock \bibinfo{title}{{The Volume Library}}.
\newblock
\newblock
\urldef\tempurl%
\url{http://schorsch.efi.fh-nuernberg.de/data/volume/}
\showURL{%
\tempurl}


\bibitem[\protect\citeauthoryear{{Schulte zu Berge}, Baust, Kapoor, and
  Navab}{{Schulte zu Berge} et~al\mbox{.}}{2014}]%
        {SchultezuBerge2014a}
\bibfield{author}{\bibinfo{person}{C {Schulte zu Berge}}, \bibinfo{person}{M
  Baust}, \bibinfo{person}{A Kapoor}, {and} \bibinfo{person}{N Navab}.}
  \bibinfo{year}{2014}\natexlab{}.
\newblock \showarticletitle{{Predicate-Based Focus-and-Context Visualization
  for 3D Ultrasound}}.
\newblock \bibinfo{journal}{\emph{IEEE Trans Vis Comput Graph}}
  \bibinfo{volume}{20}, \bibinfo{number}{12} (\bibinfo{year}{2014}),
  \bibinfo{pages}{2379--2387}.
\newblock
\showISBNx{1941-0506 (Electronic) 1077-2626 (Linking)}
\urldef\tempurl%
\url{https://doi.org/10.1109/TVCG.2014.2346317}
\showDOI{\tempurl}


\bibitem[\protect\citeauthoryear{Sitzmann, Zollh{\"{o}}fer, and
  Wetzstein}{Sitzmann et~al\mbox{.}}{2019}]%
        {Sitzmann2019}
\bibfield{author}{\bibinfo{person}{Vincent Sitzmann}, \bibinfo{person}{Michael
  Zollh{\"{o}}fer}, {and} \bibinfo{person}{Gordon Wetzstein}.}
  \bibinfo{year}{2019}\natexlab{}.
\newblock \showarticletitle{{Scene Representation Networks: Continuous
  3D-Structure-Aware Neural Scene Representations}}.
\newblock \bibinfo{journal}{\emph{Conference on Neural Information Processing
  Systems}} (\bibinfo{year}{2019}), \bibinfo{pages}{1121--1132}.
\newblock


\bibitem[\protect\citeauthoryear{Soundararajan and Schultz}{Soundararajan and
  Schultz}{2015}]%
        {Soundararajan2015}
\bibfield{author}{\bibinfo{person}{K.~P. Soundararajan} {and}
  \bibinfo{person}{T. Schultz}.} \bibinfo{year}{2015}\natexlab{}.
\newblock \showarticletitle{{Learning Probabilistic Transfer Functions: A
  Comparative Study of Classifiers}}.
\newblock \bibinfo{journal}{\emph{Computer Graphics Forum}}
  \bibinfo{volume}{34}, \bibinfo{number}{3} (\bibinfo{year}{2015}),
  \bibinfo{pages}{111--120}.
\newblock
\showISSN{14678659}
\urldef\tempurl%
\url{https://doi.org/10.1111/cgf.12623}
\showDOI{\tempurl}


\bibitem[\protect\citeauthoryear{Tewari, Fried, Thies, Sitzmann, Lombardi,
  Sunkavalli, Martin-Brualla, Simon, Saragih, Nießner, Pandey, Fanello,
  Wetzstein, Zhu, Theobalt, Agrawala, Shechtman, Goldman, and
  Zollhöfer}{Tewari et~al\mbox{.}}{2020}]%
        {Tewari2020}
\bibfield{author}{\bibinfo{person}{Ayush Tewari}, \bibinfo{person}{Ohad Fried},
  \bibinfo{person}{Justus Thies}, \bibinfo{person}{Vincent Sitzmann},
  \bibinfo{person}{Stephen Lombardi}, \bibinfo{person}{Kalyan Sunkavalli},
  \bibinfo{person}{Ricardo Martin-Brualla}, \bibinfo{person}{Tomas Simon},
  \bibinfo{person}{Jason Saragih}, \bibinfo{person}{Matthias Nießner},
  \bibinfo{person}{Rohit Pandey}, \bibinfo{person}{Sean Fanello},
  \bibinfo{person}{Gordon Wetzstein}, \bibinfo{person}{Jun-Yan Zhu},
  \bibinfo{person}{Christian Theobalt}, \bibinfo{person}{Maneesh Agrawala},
  \bibinfo{person}{Eli Shechtman}, \bibinfo{person}{Dan~B. Goldman}, {and}
  \bibinfo{person}{Michael Zollhöfer}.} \bibinfo{year}{2020}\natexlab{}.
\newblock \showarticletitle{{State of the Art on Neural Rendering}}.
\newblock \bibinfo{journal}{\emph{Computer Graphics Forum}}
  (\bibinfo{year}{2020}).
\newblock
\showISSN{1467-8659}
\urldef\tempurl%
\url{https://doi.org/10.1111/cgf.14022}
\showDOI{\tempurl}


\bibitem[\protect\citeauthoryear{Wang, Bovik, Sheikh, and Simoncelli}{Wang
  et~al\mbox{.}}{2004}]%
        {Wang2004}
\bibfield{author}{\bibinfo{person}{Zhou Wang}, \bibinfo{person}{A.C. Bovik},
  \bibinfo{person}{H.R. Sheikh}, {and} \bibinfo{person}{E.P. Simoncelli}.}
  \bibinfo{year}{2004}\natexlab{}.
\newblock \showarticletitle{{Image Quality Assessment: From Error Visibility to
  Structural Similarity}}.
\newblock \bibinfo{journal}{\emph{IEEE Transactions on Image Processing}}
  \bibinfo{volume}{13}, \bibinfo{number}{4} (\bibinfo{date}{apr}
  \bibinfo{year}{2004}), \bibinfo{pages}{600--612}.
\newblock
\showISSN{1057-7149}
\urldef\tempurl%
\url{https://doi.org/10.1109/TIP.2003.819861}
\showDOI{\tempurl}


\bibitem[\protect\citeauthoryear{Weiss, Chu, Thuerey, and Westermann}{Weiss
  et~al\mbox{.}}{2019}]%
        {Weiss2019}
\bibfield{author}{\bibinfo{person}{Sebastian Weiss}, \bibinfo{person}{Mengyu
  Chu}, \bibinfo{person}{Nils Thuerey}, {and} \bibinfo{person}{Ruediger
  Westermann}.} \bibinfo{year}{2019}\natexlab{}.
\newblock \showarticletitle{{Volumetric Isosurface Rendering with Deep
  Learning-Based Super-Resolution}}.
\newblock \bibinfo{journal}{\emph{IEEE Transactions on Visualization and
  Computer Graphics}} (\bibinfo{year}{2019}), \bibinfo{pages}{1--1}.
\newblock
\showISSN{1077-2626}
\urldef\tempurl%
\url{https://doi.org/10.1109/tvcg.2019.2956697}
\showDOI{\tempurl}
\showeprint[arxiv]{1906.06520}


\bibitem[\protect\citeauthoryear{Zhang, Zhu, Isola, Geng, Lin, Yu, and
  Efros}{Zhang et~al\mbox{.}}{2017}]%
        {Zhang2017}
\bibfield{author}{\bibinfo{person}{Richard Zhang}, \bibinfo{person}{Jun~Yan
  Zhu}, \bibinfo{person}{Phillip Isola}, \bibinfo{person}{Xinyang Geng},
  \bibinfo{person}{Angela~S Lin}, \bibinfo{person}{Tianhe Yu}, {and}
  \bibinfo{person}{Alexei~A Efros}.} \bibinfo{year}{2017}\natexlab{}.
\newblock \showarticletitle{{Real-time user-guided image colorization with
  learned deep priors}}. In \bibinfo{booktitle}{\emph{ACM Transactions on
  Graphics}}, Vol.~\bibinfo{volume}{36}.
\newblock
\showISSN{15577368}
\urldef\tempurl%
\url{https://doi.org/10.1145/3072959.3073703}
\showDOI{\tempurl}
\showeprint[arxiv]{1705.02999v1}


\end{thebibliography}

\end{document}